\input phyzzx
\hsize=40pc
\font\sf=cmss10                    

\def\Figure#1#2{\midinsert
$$\BoxedEPSF{#1}$$
\noindent {\sf #2}
\endinsert}

\input BoxedEPS
\SetRokickiEPSFSpecial 
\HideDisplacementBoxes
\catcode`\@=11 
\def\NEWrefmark#1{\step@ver{{\;#1}}}
\catcode`\@=12 

\def\footstrut{\baselineskip 12pt}
\def\square{\kern1pt\vbox{\hrule height 1.2pt\hbox{\vrule width 1.2pt\hskip 3pt
   \vbox{\vskip 6pt}\hskip 3pt\vrule width 0.6pt}\hrule height 0.6pt}\kern1pt}

\def\bra#1{\langle #1 |}
\def\ket#1{| #1 \rangle}

\def\ov{{\overline}}

\def\A{{\cal A}}
\def\B{{\cal B}}

\def\D{{\cal D}}

\def\F{{\cal F}}
\def\I{{\cal I}}
\def\K{{\cal K}}

\def\M{{\cal M}}

\def\O{{\cal O}}

\def\T{{\cal T}}

\def\V{{\cal V}}

\def\p{\partial}

\def\k{{\kappa}}
\def\l{{\bigl[}}
\def\r{{\bigr]}}
\def\ov{\overline}

\def\B{{\cal B}}
\def\D{{\cal D}}

\def\V{{\cal V}}
\def\O{{\cal O}}

\def\p{\partial}

\singlespace

\def\define#1#2\par{\def#1{\Ref#1{#2}\edef#1{\noexpand\refmark{#1}}}}
\def\con#1#2\noc{\let\?=\Ref\let\<=\refmark\let\Ref=\REFS
         \let\refmark=\undefined#1\let\Ref=\REFSCON#2
         \let\Ref=\?\let\refmark=\<\refsend}

\let\refmark=\NEWrefmark

\define\kakukikkawa{M. Kaku and K. Kikkawa, ``Field theory of relativistic
strings. I. Trees'', Phys. Rev. {\bf D10} (1974) 1110; ``Field theory
of relativistic strings. II. Loops and Pomerons'', Phys. Rev. {\bf D10}
(1974) 1823.}

\define\zwiebachhms{B. Zwiebach, `New moduli spaces from string
background independence consistency conditions', MIT preprint MIT-CTP-2527,
hep-th/9605075}

\define\zwiebachlong{B. Zwiebach, `Closed string field theory: Quantum
action and the Batalin-Vilkovisky master equation', Nucl. Phys {\bf B390}
(1993) 33, hep-th/9205075.}

\define\senzwiebach{A.~Sen and B.~Zwiebach, `Local background
independence of classical closed string field theory', Nucl.
Phys.{\bf B414} (1994) 649, hep-th/9307088.}

\define\senzwiebachtwo{A.~Sen and B.~Zwiebach, `Quantum background
independence of closed string field theory',  Nucl. Phys.{\bf
B423} (1994) 580, hep-th/9311009.}

\define\zwiebachweyl{B.~Zwiebach, `Generalized BRST operator and Weyl-descent
equations', MIT-CTP-2533, to appear.}

\define\bergmanzwiebach{O.Bergman and B.Zwiebach `The dilaton theorem
and closed string backgrounds' Nucl.~Phys. {\bf B441} (1995) 76,
hep-th/9411047.}

\define\wittenosbi{E. Witten, ``On background independent open-string 
field theory'' Phys. Rev. {\bf D46} (1992) 5467; hep-th/9208027; ``Some
computations in background independent open-string field theory'', 
Phys. Rev. {\bf D47} (1993) 3405; hep-th/9210065;\hfill\break
K.~Li and E. Witten, ``Role of short distance behavior in off-shell open-string
field theory'', Phys. Rev. {\bf D48} (1993) 853; hep-th/9303067.}

\define\zwiebachos{B. Zwiebach, ``Interpolating string field theories",
Mod. Phys. Lett. {\bf A7} (1992) 1079, hep-th/9202015.}

\define\shatashvili{S. Shatashvili, ``On the problems with background 
independence in string theory'', hep-th/9311177.}

\singlespace
{}~ \hfill \vbox{\hbox{MIT-CTP-2531}
 }\break
\title{BUILDING STRING FIELD THEORY AROUND}
\titlestyle{NON-CONFORMAL BACKGROUNDS}
\author{Barton Zwiebach \foot{E-mail address: zwiebach@irene.mit.edu
\hfill\break Supported in part by D.O.E.
cooperative agreement DE-FC02-94ER40818.}}
\address{Center for Theoretical Physics,\break
Laboratory of Nuclear Science\break
and Department of Physics\break
Massachusetts Institute of Technology\break
Cambridge, Massachusetts 02139, U.S.A.}

\abstract 
{The main limitations of string field theory 
arise because its present formulation requires a 
background representing a classical solution, 
a background defined by a strictly conformally invariant theory.
Here we sketch a construction for a gauge-invariant
string field action around non-conformal backgrounds. 
The construction makes no reference to any conformal theory.
Its two-dimensional field-theoretic
aspect is based on a generalized BRST operator 
satisfying a set of Weyl descent equations. Its geometric aspect
uses a complex of moduli
spaces of two-dimensional Riemannian manifolds having ordinary punctures, 
and organized by the number of special punctures which goes from zero 
to infinity. In this complex there is a 
Batalin-Vilkovisky algebra that includes  naturally the operator
which adds one special puncture. We obtain 
a classical field equation that appears to relax the condition of
conformal invariance usually taken to define classical string backgrounds.}
\endpage

\singlespace
\baselineskip=18pt

\chapter{Introduction and Summary}

Light-cone string field theory, 
developed in the early seventies, 
was explicitly constructed around 
Minkowski spacetime background [\kakukikkawa]. Its framework,
however, can be used to build string theory around spacetime backgrounds 
which include at least two flat dimensions, one of which is time. 
Covariant string field theory, developed from the middle eighties to the 
early nineties, removed this restriction. String field theory
can now be formulated around general conformal backgrounds, backgrounds 
encoded by strictly conformal invariant field theories (CFT's) and representing
rather arbitrary classical solutions of string theory. 
Even though this formulation does not give an {\it a priori} 
characterization of classical solutions, CFT's are clearly classical 
solutions because the 
corresponding string field action 
is quadratic in the fluctuating field. 
In passing from light-cone to covariant string field theory
much was learned about the properties and significance of the BRST operator,
the construction of covers of moduli spaces of Riemann surfaces, and the
role of Batalin-Vilkovisky algebras. 
The next logical step would be to achieve a construction of string field
theory around completely general backgrounds, backgrounds 
that need not represent
classical solutions. The resulting string field theory would have terms 
linear in the fluctuating field.  It is the purpose of the present work 
to begin such construction.

Indeed the main limitations of
 present day string field theory are not due to the
use of a `background' to write the theory. The limitations largely 
arise because 
the background must correspond to a classical solution. Ideally
string field theory should help us find classical solutions, but
finding classical solutions using a reference classical solution is
not very easy.
Moreover, it is not manifest that string field theories formulated using
different conformal backgrounds are really the same theory. This is the
question of background independence, a property that was proven explicitly 
in Refs.[\senzwiebach, \senzwiebachtwo].   One may wonder if
a fully satisfactory string
action should be written without using a background at all, but
comparison with Einstein's theory, a familiar and natural analog, suggests
otherwise.  
In gravity, backgrounds play the role of off-shell
fields and represent arbitrary metrics on a manifold.  
The action is  a function in the space of 
backgrounds, a function that is stationary at the backgrounds 
representing classical solutions.  Background independence 
 for Einstein's theory is simply
the fact that the space of backgrounds (the space of metrics) can
be  described without using a special background (a special metric).
It is natural to wonder  what is the space of string backgrounds.
Present day string field theory provides a partial answer:  the
space of string backgrounds in the neighborhood of a conformal background
is the vector space of all local operators of the conformal theory.\foot{
Restricted to operators annihilated by $L_0^-$ and $b_0^-$, in
the case of closed strings.}  It has not proven easy to give a description
of the space of backgrounds without using a reference conformal background, 
but the expected answer has
been that this space is something closely related
to the space of two-dimensional field theories. 

Indeed, we  say something {\it closely related~}
to the space of two-dimensional field theories, since subtle
complications abound. The field theories we must consider must have 
all possible couplings, including non-renormalizable ones.
This may not be a serious difficulty, if all couplings are present
non-renormalizability is really not an issue, the space of theories
is just the space of renormalized couplings, a space of infinite dimension.
Another issue, possibly more relevant [\bergmanzwiebach], 
is the recent finding that the
space of string backgrounds have parameters that are not present in the
space of quantum field theories. This is the case of the string coupling.
The string coupling arises from the expectation value of the string dilaton and
is a parameter of string backgrounds but it is not a parameter of the
conformal theories (due to the ghost anomaly). This suggests
that in some sense the space of backgrounds is a little more akin to the
space of possible two-dimensional Lagrangians.

In this paper the ingredient we will use for the construction of 
the string action is  a two-dimensional 
quantum field theory, a non-conformal one.
How close it is to a general two-dimensional quantum 
field theory is something that one may  learn by further study of the
construction to be given here.  
The  string action will be written as a function on
the state space of this two-dimensional quantum 
field theory. 
This state space  represents the tangent space to the space  
of backgrounds
at the particular background encoded by the field theory in question. 
As such, the state space can only
be expected to give a local description of the space of backgrounds.
Ideally we would like to have 
the string action defined directly on the space of backgrounds.
Strictly speaking, the  construction to be presented here
still has a residual background dependence since the string action is not
written directly on the space of backgrounds but rather on its tangent space
at some background: the identification of the tangent
space with the space itself can be background dependent.
Thus given two backgrounds, the string
actions written in the two different tangent spaces may not be manifestly
the same. It seems very plausible that a background independence analysis would
enable one to rewrite the action to be built in this paper as a function on the
space of backgrounds.   

There are two pieces of indirect evidence that suggest that the
above construction of a string field theory around non conformal
backgrounds should be possible. 
First, by giving
an arbitrary expectation value to the string field of a SFT formulated around
a conformal background we obtain a well-defined string action which is
not stationary in the new fluctuating field [\zwiebachlong]. This
action is gauge invariant and represents string theory around a background
which is not a classical solution. 
The reason this action is not the desired answer is that
the new background is described explicitly in terms of the
original conformal background.
The second piece of evidence comes from the sigma model approach 
to string theory. In this approach given 
fairly arbitrary two-dimensional field theories 
one can define beta functions whose vanishing appear
to represent string field equations. 
In this approach one does not assume conformal
invariance. The sigma model method suffers from several technical 
complications: it is very difficult to deal with backgrounds that
correspond to massive string fields,
the definition of the list of background fields is not systematic and, 
there is no prescription to find the action, nor
its gauge invariances. 

In the standard formulation of string field theory gauge transformations
are defined by the BRST operator, and gauge invariance is a consequence
of the nilpotency of this operator. It should be emphasized that
string field theory formulated around non-conformal backgrounds must
still be gauge invariant, despite the fact that one will not be able
to have a BRST operator that squares to zero.
Earlier work trying to write a string action 
as a function in the space of
two dimensional field theories has been mostly inconclusive. 
The functions that were written do not
show any clear evidence of string field gauge invariances. 
\foot{With the notable
exception of Ref.[\wittenosbi] which addressed the case of open
strings using the Batalin-Vilkovisky framework
to ensure gauge invariance from the start. Difficulties with this approach 
have been studied in Ref.[\shatashvili].}
While the condition to have a conformal background is the vanishing of 
the trace
$T\equiv T_\mu^\mu$ of the stress tensor, it is not clear that the equation 
of motion that selects a {\it classical string background}
is simply $T =0$. Such equation of motion need not have string 
field gauge invariance.
In fact we will find that the equation of motion of
string theory  appears to be
non polynomial in $T$, with leading term linear in $T$. 
Vanishing $T$
would  always be a solution, but there
might exist solutions with non zero $T$, such solutions
would represent classical string backgrounds that are not conformal. 

The construction of string field theory around a non conformal background will
have two ingredients, one from two-dimensional geometry and the other from
two-dimensional field theory. Both ingredients were also present in the case
of conformal backgrounds but as we will see there are major departures.
Let us first consider the two dimensional geometry ingredient. 

\goodbreak
\noindent
\underbar{Moduli spaces of two-dimensional surfaces}.~In the conformal case
we found string vertices, collectively denoted by $\V$ which represented
pieces of the moduli spaces of Riemann surfaces with punctures and 
local coordinates at the
punctures. The regions of moduli spaces corresponding to the string vertices,
and the local coordinates at the punctures had to be chosen  carefully
in order that the string vertices satisfy a set of recursion relations involving
sewing. Such choices were made using an auxiliary problem on Riemann surfaces,
that of finding a metric (a Weyl factor) of least area under the
condition that all nontrivial closed curves be longer than or equal to $2\pi$.
The local coordinates at the punctures could be defined using the behavior
of the metric near the punctures. While the metric played a purely auxiliary
role in the conformal case, in the non conformal case it is necessary
since the two-dimensional field theory is not 
conformal and  correlation functions will depend on the chosen metric. 
Since the recursion relations hold when the string vertices are
equipped with metrics and local coordinates, 
they can be used in the non conformal case. It will not be possible, however,
to attain gauge invariance using only the string vertices. Apart from
two-dimensional quantum field theory complications, 
new moduli spaces of Riemann
surfaces equipped with Weyl metrics are needed. 

The need arises from the algebraic structure of the theory to 
be built. Such structure was explored indirectly Ref.[\zwiebachlong] sect 4.5.
The algebraic structure of standard closed string field theory
is based on a homotopy Lie algebra defined by string field products
$m_n$ where $n\geq 1$ is the number of string fields to be multiplied.
The lowest product $m_1$ is simply action by the BRST operator of
the conformal theory. For the present case the homotopy Lie algebra must have
a product $m_0$. Such product simply represents a particular (fixed) 
string field, a grassmann odd, ghost-number $+3$ string field
to be denoted as $F$. The surfaces in the 
new moduli spaces that we need
will have special punctures where the string field $F$ will be inserted.
In addition the surfaces will also have ordinary punctures where the
dynamical string fields are inserted. 
The $\B^1$ spaces of background independence [\senzwiebachtwo], 
now equipped with Weyl metrics,
have one special puncture. The special string field $F$ will be inserted
at this special puncture, and the dynamical string field is inserted
at the ordinary punctures. The resulting function of the string field
will define the part of the string action linear in $F$. 
The identities satisfied by the $\B^1$ spaces that
guaranteed background independence in [\senzwiebachtwo]
 are precisely the identities that
guarantee gauge invariance in the new construction. 
With this linear term included gauge invariance will
hold to $\O( F)$, but not to quadratic order.
Moduli spaces $\B^2$ of surfaces with two special punctures are required. 
Such kind of moduli spaces were studied in detail in [\zwiebachhms] where 
they were seen to arise from second order background independence conditions. 
Again, the identities required for background independence
are essentially the same identities that are required to attain gauge
invariance in the new construction. The $\B^2$ spaces with  $F$'s inserted
in each of the two special punctures defines the part of the string action
quadratic in $F$, and gauge invariance now holds to $\O (F^2)$.
To achieve full gauge invariance we introduce moduli spaces
$\B^3, \B^4, \cdots$ with all numbers of special punctures. 
While we work with such spaces directly, we will sketch how such 
spaces also arise from higher order background
independence conditions. We thus see that the geometrical
ingredient necessary to build a string field theory around
a non-conformal background was encoded 
in the potentially infinite set of background independence conditions that
arise from a string field theory built around a background encoded by
a CFT that sits on a CFT space.   

We explain that all $\B$ spaces $\B^1, \B^2, \cdots$ should be thought as
string vertices, with the standard string vertices $\V$, which have
no special punctures, identified as the space $\B^0$. 
The complete collection of all string 
vertices can be formally put together into a single
space $\B \equiv \B^0+ \B^1 + \B^2 + \cdots$. This element $\B$ is an
element of the complex which includes the formal sum
of moduli spaces of Riemann spheres
with Weyl metrics with all possible numbers of ordinary and 
special punctures.\foot{In this paper we only consider the 
classical closed string theory and thus only spheres are relevant.
The extension to quantum closed strings is not expected to be
problematic.} We write $\B = \sum_{k,p} \B^k_p$ where $\B^k_p$ denotes
a moduli space of spheres with $k$ special punctures and $p$ ordinary
punctures.  
In this complex we have an antibracket operation, which corresponds
to the sewing of ordinary punctures of the corresponding surfaces, and
an operator $\K$ that acting on a moduli space of surfaces it adds one 
special puncture to the each of the surfaces [\zwiebachhms]. 
The operator $\K$ satisfies the identity
$\K^2=0$, and this is related to the fact that any  space  in the complex
is defined to be antisymmetric
under the exchange of the labels on any two of the special punctures.  The 
generalized string vertex $\B$ is shown to satisfy an 
extremely simple relation: $\partial\B  -\K \B +\half
\{ \B , \B \} - \V'_3 =0$. If expanded in terms of the number of special
punctures this equation gives all the recursion relations satisfied by
the familiar string vertices $\V$ and all the $\B^k$ spaces. This equation
explains in a simple way why the action we build satisfies
the Batalin-Vilkovisky master equation.

It is important to note that the label $k$ in 
$\B^k_p$ not only gives the number of special punctures on the surfaces
in the moduli
space $\B^k_p$ but also specifies the dimension of $\B^k_p$. We have:
$\hbox{dim}~(\B^k_p) = k+ \hbox{dim}~(\V_{k+p})$, namely, for each
special puncture the moduli space gains one real dimension above that
of the standard string vertex with the same total number of punctures.    
This is in accord with ghost number conservation, given that $F$ has
ghost number three, and that ordinary string fields 
giving a nonzero contribution
to an ordinary string vertex must have an average ghost number of two.
One can therefore think of $\B^k_p$ as a space fibered over $\M_{k+p}$ where
$\pi: \B^k_p \to \M_{k+p}$
is a projection that forgets about the Weyl metric on the surface and
the choice of local coordinates at the punctures. 
For every complex structure $\Sigma \in \pi \,(\B^k_p)$ 
one has a {\it compact} $k$ dimensional space $\pi^{-1} (\Sigma)\in \B^k_p$ 
with the same complex structure. 
In the conformal context the
fibers simply represented different local coordinates at the punctures.
Integration over $\B$ spaces includes 
integration over fibers, and in the present case
this is properly thought as integration
over Weyl metrics! This represents a surprising realization of 
an intuitive expectation. If we start with
non conformal theories one would expect that integration over metrics
would include integration over conformal
structures {\it and} Weyl metrics. This could not have been strictly true, 
however. A complete integration over  Weyl metrics
would very likely give rise to infinities,
since the space of inequivalent Weyl metrics 
is infinite dimensional and non-compact.
The standard string vertices have no Weyl integration, but the $\B^k$ spaces
do. Since they are compact finite dimensional spaces they give rise  to
limited integration over the space of Weyl metrics. 
As the number of special punctures
increases one is integrating over compact 
subspaces of the space of Weyl metrics of larger and larger dimensionality.

\noindent
\underbar{Two dimensional field theory ingredient.}
~The main question here is what is the string field $F$ and what
is the proper replacement of the BRST operator, which for the
case of conformal backgrounds defines the linearized gauge 
transformations of the theory. In the conformal context,  
gauge invariance, to first order,
follows from the fact that the BRST operator squares to zero.
For two-dimensional non-conformal field theories we do not
expect to have a BRST charge that is conserved nor we expect this charge
to square to zero. 

There is a way to motivate the type of identities that must be satisfied
by the generalized BRST charge $Q$ and the string field $F$. 
The type of identities required for gauge invariance 
away from conformal backgrounds were given in Ref.[\zwiebachlong]
sect.4.5, and the first few read 
$$  {\cal Q}\ket{{\cal F}} = 0\, , \eqn\clfornzero$$
$${\cal Q}^2\, \ket{A} = -  \l\, {\cal F} , A \r \, ,\eqn\failnilp$$
$$ {\cal Q} \l A_1 , A_2 \r  + \l {\cal Q} A_1 , A_2 \r + (-)^{A_1} \l A_1,
{\cal Q}A_2 \r  = -  \l {\cal F},A_1,A_2 \r  , \eqn\foxxfxo$$
We have used calligraphic symbols to denote ${\cal Q}$ and ${\cal F}$ because
these are not the same objects 
as $Q$ and $F$. We need ${\cal Q }$ and $\F$ but it seems  unlikely that 
one can give
a simple construction of them starting from a two-dimensional quantum
field theory. Indeed, ${\cal F} =0$ is the
condition that selects a classical string background, and
such equation is probably fairly intricate. Our strategy will be to 
construct ${\cal Q}$ and ${\cal F}$ from simpler objects $Q$ and $F$
whose existence we will {\it postulate}. The object $Q$ will correspond to
a non-conserved charge and therefore will be contour
dependent and denoted as $Q(\gamma)$. We demand that
$$\eqalign{
\lim_{r\to 0} Q (\gamma_r)\, F (0)\, &=\,\,\,0 \,\,\,,\cr
[ Q (\gamma) ]^2\, &=\,\, {1\over 2\pi i}\, \int_\gamma F^{\,[1]} \,,\cr
Q(\gamma_2) - Q (\gamma_1)\, &= - {1\over 2\pi i} \int_M F^{\,[2]} \,. \cr}
\eqn\conse$$ 
We claim that these equations define a natural extension 
of the conformal field theory BRST operator to the non conformal case. 
The special ghost
number three local operator $F(z,\bar z)$, vanishing in the conformal
case, controls the violation of all the standard BRST properties.    
The first equation says that as we shrink the contour around $F$ the
BRST operator will give a result that goes to zero.
This is the closest analog of \clfornzero\  given that the BRST operator is not
conserved. The next identity equates the failure
of $Q(\gamma)$ to square to zero to the line integral of
the one-form $F^{\,[1]}(z,\bar z)$ associated to $F (z, \bar z)$.
Finally the failure of $Q$ to be conserved is measured by the variation
of $Q$ as we change the defining contours. For 
two homologous contours $\gamma_1$ and $\gamma_2$,
namely two contours bounding a cylindrical region $M$: $\partial M =\gamma_2
-\gamma_1$ we find that the change in $Q$ is given by an integral of the
two-form $F^{\,[2]}(z, \bar z)$ associated to $F$. 

We will not
attempt in the present paper to give an explicit construction of the
operators $Q$ and $F$ in terms of two-dimensional field theory data.
This will be the subject of a forthcoming publication [\zwiebachweyl].
The BRST operator will be built in terms of line integrals involving
ghost fields and the stress tensor of the theory (which is not traceless).
The two form $F^{[2]}$ appears to be equal to the divergence of the
ghost field times the trace $T$ of the stress tensor. Apart from states
that have nontrivial ghost dependence, the operator $F$ is also proportional
to the trace $T$ of the stress tensor. The ghost-sector of the two-dimensional
field theory must be defined so that the BRST operator acts properly
on forms on the moduli space of Riemann surfaces equipped with Weyl metrics.

The relation between $Q$ and $F$ on one hand and ${\cal Q}$ and ${\cal F}$ on
the other is through the family of $\B$ spaces discussed earlier. In fact
${\cal Q}$ can be identified as the operator appearing in the part of the
string action quadratic on the string field and ${\cal F}$ as the string 
field appearing in the part of the action linear on the string field.
In the case of ${\cal F}$ it turns out that it equals $F$ plus contributions
from $\B$ spaces with one ordinary puncture and two or more special punctures,
namely $\B^k_1$ spaces with $k\geq 2$. The complete field equation reads:
$$\ket{{\cal F}}_{1'} \equiv 
 \ket{F}_{1'} +  \sum_{k\geq 2}\,{1\over k!} \int_{\B^k_1} 
\bra{\Omega^{[k]}_{1\bar 1 \cdots \bar k}}
F\rangle_{\bar 1}\cdots \ket{F}_{\bar k} \ket{S_{11'}}  =0\, . \eqn\feqn$$
This is a non polynomial equation that starts with a term linear in $\ket{F}$
and by construction $\ket{F} =0$ is a solution. As we mentioned earlier 
there may be nontrivial solutions of this equation. Such solutions, having
$\ket{F}\not= 0$, but $\ket{{\cal F}}=0$ would represent classical 
string backgrounds
that do not correspond to conformal field theories.

\Figure{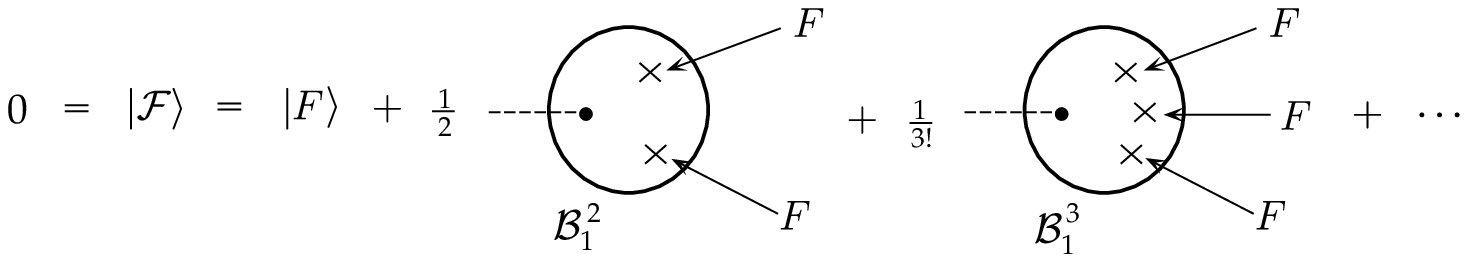}{Figure 1.~The representation of the field equation 
that selects a classical string background. The first term in the 
equation is simply the state $\ket{F}$, the second term is the state
obtained through the ordinary puncture in the moduli 
space $\B^2_1$ where the two special punctures are filled with $F$. }

It is worthwhile to point out briefly the parallel with 
Einstein's theory. The
utility of Einstein's equations derives mostly from the fact that they can
be written without committing one-self to a reference classical background.
The equations of motion for the fluctuating field around flat space would
be of little help in finding nontrivial classical solutions.
The Einstein action, expanded around an arbitrary
background metric $\ov g_{\alpha\beta}$ that does not satisfy the
field equations reads
$$ S(g) = -{1\over \k^2} \int \sqrt{\ov g}\, \ov R\, d^4x
+{1\over \k^2} \int d^4x \sqrt{\ov g}\, \l\, \ov R^{\mu\nu}
- {1\over 2} \,\ov g^{\mu\nu}\, \ov R\, \r \, h_{\mu\nu}
+ \O (h^2), \eqn\einst$$
where $h_{\mu\nu}$ is the fluctuating field. 
The advantage of expanding around an arbitrary metric is clear. Multiplying
the fluctuating field is the fully geometrical Einstein field 
equation:
$\ov R^{\mu\nu} - \half\,\ov g^{\mu\nu}\, \ov R\, =0$. Our string analog
for this equation is $\ket{{\cal F}(F) }=0$, as shown in \feqn. 
Gauge invariance of \einst\ under
$\delta_\epsilon h_{\mu\nu} = \ov D_\mu \epsilon_\nu+ \ov D_\nu\epsilon_\mu$
holds on account of the Bianchi identity
$\overline D_\nu \l \,\ov R^{\mu\nu} - \half\,
\ov g^{\mu\nu}\, \ov R\, \r = 0$.  The string field theory analog 
of the gravitational Bianchi identity is
the equation ${\cal Q }\ket{{\cal F}} =0$. The field independent term
in the expansion \einst\ is simply the Einstein action evaluated at
the background metric. It is for this reason that it should be of
interest to understand the $\B^k_0$ spaces. These spaces
give the string field-independent contributions to the string field action
(see conclusions, for further comments).

\noindent
\underbar{Organization of this paper}~
In section 2 we first review a few of the results of Ref.[\zwiebachhms] 
and then discuss the systematics of
 higher order background independence conditions, showing how they arise
from the requirement of vanishing antibracket cohomology. 
In section 3 we explain the identities that should be satisfied by the
generalized BRST operator $Q$ and the special string field $F$. We explore
the consistency of these equations, and discuss $Q$ action on moduli
spaces whose surfaces have both ordinary and special punctures. In
section 4 we begin the construction of the new gauge invariant string 
action as a power expansion in the number of $F$ insertions. 
We discuss explicitly the terms with zero, one and two insertions. Such
terms use the string vertices $\V$, the $\B^1$ spaces and $\B^2$ spaces.
In section 6 we complete the construction by working to all orders
in the number of $F$ insertions. We offer comments and spell out
some open questions in section 7.

\chapter{Review and Developments}

In this section we review some of the technology developed in [\zwiebachhms]
dealing with moduli spaces of surfaces with ordinary and special punctures.
While the cases of one or two special punctures were the subject of
main attention in previous works, in the present paper the number of
special punctures will be arbitrary. We  define in this general context
the homomorphism from surfaces to functions. Finally, we explain how
relevant $\B$ spaces with more than two punctures arise from higher order
background independence consistency conditions.

\section{Properties of String Vertices and $\B$ spaces}

The moduli spaces of surfaces we must deal with are moduli 
spaces of Riemann surfaces with labelled punctures. The punctures can
be of two types, ordinary punctures, where the dynamical string field is
inserted, and special punctures where special states are inserted.
The punctures have analytic local coordinates defined around them,
and the moduli spaces are symmetric under the exchange of labels of
any two ordinary punctures, and antisymmetric under the exchange of 
labels of any two special punctures. For the purposes of the present
paper one must also define a Weyl metric (or conformal factor $\rho$
with $ds = \rho |dz|$) on every surface of each moduli space. 
The above moduli spaces, called $\B$ spaces, are indexed by the number
of special punctures, sometimes indicated as a superscript. The Batalin
Vilkovisky algebra of surfaces with ordinary punctures
[\senzwiebachtwo], was extended to this more general case in Ref.[\zwiebachhms].
The main results are summarized below. 

The antibracket of $\B$ spaces satisfies the following identities 
[\zwiebachhms]:
$$\eqalign{
\{\B_1 , \B_2\} &= -\,(-)^{(\B_1+\bar n_1 +1)(\B_2+\bar n_2 + 1)} 
\,\{\B_2,\B_1\}\,, \cr
 \p\, \{\B_1 ,  \B_2\}  &= \{ \p\B_1 ,  \B_2\} + (-)^{\B_1 +\bar n_1 +1} 
\{ \B_1 , \p\B_2 \} \,, \cr 
0 & =  (-)^{(\B_1+\bar n_1 + 1) (\B_3+\bar n_3 + 1)} \{ \{ \B_1, B_2 \} , \B_3\} 
  \, + \, \hbox{Cycl}\,. \cr}\eqn\jacobin$$
These identities work as if $\B$ spaces with $\bar n$ special punctures had an
effective dimensionality equal to  dim($\B$) + $\bar n$.
The operators $\K$ and $\I$ satisfy the relations
$$\eqalign{
\K\, (\,\{ \B_1 , \B_2\}\, )\, 
&=\,(-)^{\B_2 + \bar n_2 + 1} 
\{ \K\B_1 ,\B_2\}
 \,+\,\{ \B_1 , \, \K\B_2 \} \,,\cr
[\,\partial\,,\,\K\,]\, \B &= \,(-)^{ \B + \bar n}
\{\,\V'_3 \,, \B\, \}\,,\cr
\K ~\K & = 0\,\,, \cr} \eqn\knewd$$
and,
$$\eqalign{
\I\, (\,\{ \B_1 , \B_2\}\, )\, 
&=\,(-)^{\B_2 +\bar n_2 + 1} 
\{ \I\B_1 ,\B_2\}
 \,+\,\{ \B_1 , \, \I\B_2 \} \,,\cr
[\,\partial\,,\,\I\,]\, &=\,0 \,,\cr
\I ~\I & = 0\,\,. \cr}\eqn\iinst$$
The anticommutator of $\K$ and $\I$ is given by 
$$ \K~ \I + \I ~\K 
= \{ \,\,\,  , \T^2_1\,\,\}\,\,,\eqn\import$$   
where 
$\T^2_1 = \tau^2_1(0)  - P \tau^2_1(0)$.
Here $P$ is the operator that exchanges the labels of the special punctures,
and  $\tau_1^2(0)$ is a collection of three punctured spheres 
$\tau_1^2(0) = \Bigl\{ S(t)\,  \bigl|\,  t \in [0,1]  \Bigr\}\,$ (see Figure 1,
and Ref.[\zwiebachhms] for further details).
This space satisfies 
$$\partial \T^2_1 =  \I \V'_3 \,.\eqn\boundtau$$ 

\Figure{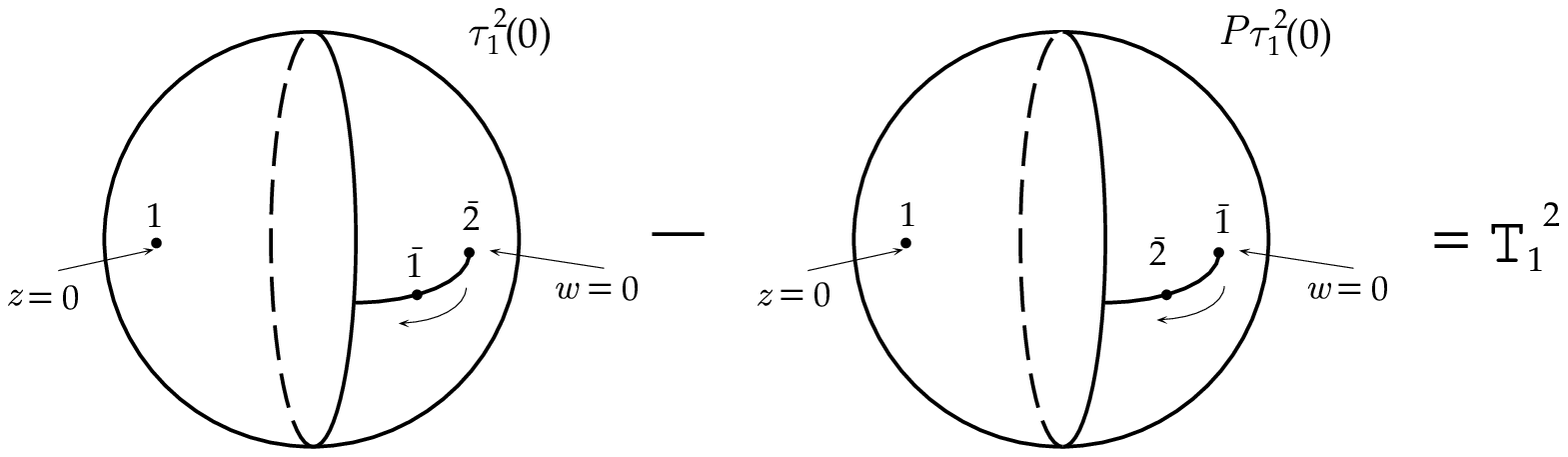}{Figure 2.~The one dimensional moduli space 
of three punctured spheres that defines the
space $\T^2_1$ having two special punctures and one ordinary puncture. 
The first space is $\tau^2_1(0)$ and the second one is $P\tau^2_1(0)$,the same
space with the labels for the special punctures exchanged. One of the special punctures travels from the point 
$w=0$, where it coincides with the other special puncture, up to the
point $w=1$.}

Defining $\M = \K - \I$ one has the following identities
$$\eqalign{
\M\, (\,\{ \B_1 , \B_2\}\, )\, &=\,(-)^{\B_2 + \bar n_2+1} 
\{ \M\B_1 ,\B_2\}
 \,+\,\{ \B_1 , \, \M\B_2 \} \,,\cr
[\,\partial\,,\,\M\,]\,\B &=\,(-)^{\B + \bar n}
\{\,\V'_3\,, \B\,  \}\,,\cr
\M ~\M & = - \, \{ \,\,\,  , \T^2_1\,\}\, .\cr}\eqn\mopiden$$

\goodbreak
\noindent
\underbar{$\V, \B^1,\, \hbox{and}\,\,\B^2$ Spaces}.~
At genus zero, the string vertices $\V$ 
include moduli spaces of punctured spheres with three or more punctures.
They satisfy the identities
$$\partial \V +  \, \half\{ \V ,\V\} =0\,. \eqn\recrel$$
$$\delta_\V^2 =0\, ,\quad \hbox{with}\quad
\delta_\V \equiv \partial + \{  \V\, \, ,\,\, \} \eqn\anuop$$
$$ \delta_\V \V = \half \{ \V \,, \, \V \} \,. \eqn\neweq$$
As defined in [\senzwiebachtwo], genus zero $\B^1$ spaces begin with a moduli
space of three punctured
spheres, one of which is a special puncture. The moduli spaces $\B^2$
at genus zero also begin with a moduli
space of three punctured
spheres, but this time two of the punctures are special.
Their recursion relations take the form [\senzwiebachtwo,\zwiebachhms]:
$$\eqalign{\delta_\V\B^1\, &= \V'_3 + \M\V
\,, \cr
 \delta_\V  \B^2  &= \T^2_1 +  \M \,\B^1  -\, \half\,
 \{ \B^1 \,,\, \B^1 \} \,. \cr} \eqn\gmallow$$
In proving the consistency of the above equations one uses the following
equalities, valid for arbitrary $\B$ spaces
$$\eqalign{
[\, \,\delta_\V ,\K\, ]\B &= \, (-)^{\B+ \bar n}
\bigl\{ \, \V'_3 + \K\V \, ,  \B\, \bigr\}\,, \cr 
[\,\,\delta_\V ,\I\, ]\B &= \, (-)^{\B+\bar n} \bigl\{  \,\I\V\,, \B\, \bigr\}
\,,    \cr
[\,\delta_\V ,\M\, ]\B &= \,  (-)^{\B+ \bar n}\bigl\{\, \V'_3 + \M\V\,,\B\, 
\bigr\}\,. \cr} \eqn\fcomb$$

The spaces $\V, \B^1,\B^2$, and the higher spaces $\B^k$ we will
introduce later have something in common.
Their dimensionality exceeds
that of the corresponding moduli space of Riemann surfaces by the
number of special punctures. As a consequence they
are effectively Grassmann even. They will all be thought eventually
as string vertices.
It follows  from the results listed earlier
that they satisfy the following identities
$$ \bigl\{ \,\B^{k_1} \,,\, \B^{k_2} \, \bigr\}
= \bigl\{ \,\B^{k_2} \,,\, \B^{k_1} \, \bigr\} \,,\eqn\nbmv$$  
$$ \M \bigl\{ \,\B^{k_1} \,,\, \B^{k_2} \, \bigr\}
= - \bigl\{ \,\M\B^{k_1} \,,\, \B^{k_2} \, \bigr\} 
+ \bigl\{ \,\B^{k_1} \,,\, \M\B^{k_2} \, \bigr\}\,, \eqn\lop$$
$$ \Bigl\{ \{ \B^{k_1}\, , \,   \B^{k_2} \} \, , \,  \B^{k_3}\Bigr\} + 
\hbox{Cyclic} = 0 \,. \eqn\newjac$$

\section{Mapping to string functionals}

Throughout this paper we 
will insert the ghost number three, Grassmann odd $F$ states at every special 
puncture. It is therefore convenient to define 
$$f(\B^k_n)   = {1\over n!} {1\over k!} \int_{\B^k_n}
\bra{\Omega} \Psi\rangle_1 \cdots \ket{\Psi}_n \ket{F}_{\bar 1}
\cdots \ket{F}_{\bar k}\, . \eqn\homdef$$
Note that we have included a symmetry factor both for the
ordinary punctures and for the special punctures.
Moreover,  for string vertices $\B^k_n$, the function
$f(\B^k_n)$ is Grassmann even, as it should be in order to 
be a candidate for a term in the string action.  The subscript on
each $F$ denotes the label of the puncture where the state is inserted.
Note that the ordering of the $F$ states, ascending from left to right,
is important since the $F$ states are Grassmann odd.

The main homomorphism identity arises as we consider the antibracket
of two functions arising from $\B$ spaces. Using the primed
antibracket of Ref.[\zwiebachhms], we find
$$\bigl\{ \,f (\B^{k_1}) \,,\, f (\B^{k_2} )\, \bigr\} =  - 
(-)^{k_1(k_2+1)} f\Bigl( \bigl\{ \,\B^{k_1} \,,\, \B^{k_2} \, \bigr\}'
\Bigr)\,,\eqn\gtns$$ 
where the sign factor requires careful consideration. 
\foot{It arises as follows: there is one
minus sign from the sign factor in the right hand side of Eqn.(2.10) of
Ref.[\zwiebachhms]. There are ($k_1k_2 + k_1 + k_2$) minus signs
that arise from moving all the $F$ states to 
the right. Finally there are $k_2$ minus signs from moving the 
sewing ket $\ket{S}$ in between the  operator valued forms representing 
the moduli spaces, as in Ref.[\senzwiebachtwo] Eqn.(3.12).} 
The normalization works out correctly as a consequence of the inclusion
of the factor $(1/k!)$ in the definition \homdef.
As a consistency check notice that both the rhs and the lhs, are symmetric
under the exchange $k_1 \leftrightarrow k_2$. 
Recalling the relation between the primed and unprimed antibracket
$\{ \B^{k_1} ,\B^{k_2} \} \equiv (-)^{k_1 + k_1k_2} \{  \B^{k_1} ,\B^{k_2} \}'$
we find
$$\bigl\{ \,f (\B^{k_1}) \,,\, f (\B^{k_2} )\, \bigr\} =  - 
 f\Bigl( \bigl\{ \,\B^{k_1} \,,\, \B^{k_2} \, \bigr\}
\Bigr)\,.\eqn\gtnss$$
This equation is valid for arbitrary $\B$ spaces. 
Defining now 
$$B^{(2)}_{\O}\equiv \bra{\omega_{12}\,}\O\rangle_1\ket{\Psi}_2\,\,,\eqn\lham$$
we obtain another useful homomorphism identity is 
$$f(\I \B^k) = \{ f (\B^k) , B_F^{(2)} \}\,.  \eqn\iopin$$

\section{Vanishing Antibracket cohomology classes}

Starting from the master equation $\{ S, S\} =0$, 
 covariant differentiation with
a symplectic connection immediately yields $\{ S , D_\mu S \} =0$. This means
$D_\mu S$ is $S$-closed. We expect, however, that $D_\mu S$ is actually 
$S$-exact. If it were not so, $D_\mu S$ could be added to the string action
to define an inequivalent string action, something we do not expect
physically to be true. Therefore, there should be a $B_\mu$ such that
$$ D_\mu  S = \{ S\, , \, B_\mu \} \,.\eqn\one$$
Finding the explicit form of $B_\mu$ was the subject of Ref.[\senzwiebach].
This last equation can be written briefly by introducing a
covariant
derivative acting on functions on the vector bundle [\zwiebachhms]
$$\D_\mu S = 0 \,, \quad \D_\mu \equiv D_\mu  + \{ B_\mu,\,\,\} \,.\eqn\nnot$$
Note that this covariant derivative acts nicely on the antibracket:
$\D_\mu \{A , B\} = \{\D_\mu A , B\}+ \{A , \D_\mu B\}$.
Recalling that $[D_\mu , D_\nu]\, A = -\,\{ A \, , R_{\mu\nu} \}$,
we find that the commutator of two covariant derivatives gives
$$[\D_\mu , \D_\nu ]  = - \,\{ H_{\mu\nu}\, , \,\,\, \} \,, \quad
 H_{\mu\nu} \equiv  D_\mu B_\nu
 -  D_\nu B_\mu +  \{ B_\mu  ,   B_\nu \} + R_{\mu\nu} \,.\eqn\nmot$$
It follows from \nnot\ that $ [\D_\mu , \D_\nu ]\,S = 0$, and  this
together with \nmot\
implies that $H_{\mu\nu}$ is $S$-closed:
$$\bigl\{ \,\,  S\, , \, H_{\mu\nu}\bigr\}=0 \,. \eqn\six$$
Since we do not expect nontrivial antibracket 
cohomology classes we are led to write
$$\quad H_{\mu\nu}  =  
\bigl\{ \,\,  S\, , \, B_{\mu\nu} \, \bigr\} \,.\eqn\sixx$$
Much of the work in [\zwiebachhms] went into
constructing the hamiltonian $B_{\mu\nu}$ and showing that it 
was defined by moduli spaces of surfaces having two special punctures.
We now show that higher conditions arise and could be used to 
define moduli spaces of surfaces having more than two special punctures.

As a first step we prove a ``Bianchi identity" for $H_{\mu\nu}$. 
Note the trivially
satisfied identity
$$ [\D_\mu\,, \D_\nu ]  \D_\rho + \hbox{Cycl}  =  
\D_\mu\  [\D_\nu \,,  \D_\rho\, ] \,  + \hbox{Cycl}\,, \eqn\cyclc$$
where `Cycl' denotes adding the two cyclic permutations of $(\mu\,\nu\rho)$
Letting both sides of the equation act on an arbitrary function $A$, one
deduces that
$$\D_\mu H_{\nu\rho} + \D_\nu H_{\rho\mu} + \D_\rho H_{\mu\nu} =0\,.\eqn\bian$$
It follows now from \six\ and \bian\ that 
$$\bigl\{ \,\,  S\, , \,H_{\mu\nu\rho}
\, \bigr\} = 0\,, \quad 
H_{\mu\nu\rho} \equiv \D_\mu B_{\nu\rho} + \D_\nu B_{\rho\mu} 
+ \D_\rho B_{\mu\nu}\,, \eqn\serf$$
where we have introduced a three-index field strength $H_{\mu\nu\rho}$.
Again, expecting no nontrivial antibracket cohomology, we are led to
write 
$$H_{\mu\nu\rho} = \{ S \,, B_{\mu\nu\rho} \, \}\,.  \eqn\erf$$
The new hamiltonian $B_{\mu\nu\rho}$ would be expected to arise from 
moduli spaces of surfaces with three special punctures.  

The idea to go to higher orders is simple, we take a covariant derivative
$\D_{\mu_0}$ of the equation 
$H_{\mu_1\cdots \mu_k} = \{ S , B_{\mu_1\cdots \mu_k} \}$, and antisymmetrize
on the $(k+1)$ indices. The left hand side is written in the form 
$\{ S , ~\cdot~ \}$ with the help of lower order identities, and a new
identity of the form $\{ S , H_{\mu_0 \cdots \mu_k} \} =0$ follows. Rather
than introducing general notation to do this efficiently we limit
ourselves to consider the next case. Defining
$$ \D_\mu H_{\nu\rho\sigma} \pm\hbox{Cycl.} 
\equiv  \D_\mu H_{\nu\rho\sigma} - \D_\nu H_{\rho\sigma\mu} 
+ \D_\rho H_{\sigma\mu\nu }  - \D_\sigma H_{\mu\nu\rho}\,, \eqn\ngh$$ 
it then follows from \serf\ and \nmot\ that
$$\eqalign{\D_\mu H_{\nu\rho\sigma} \pm\hbox{Cycl.}  
&= -\, \{ H_{\mu\nu} \,,\, B_{\rho\sigma} \} 
-\{ H_{\mu\rho} \,,\, B_{\sigma\nu} \} 
-\{ H_{\mu\sigma} \,,\, B_{\nu\rho} \}  \cr 
& \quad  -\, \{ H_{\rho\nu} \,,\, B_{\sigma\mu} \} 
-\{ H_{\sigma\nu} \,,\, B_{\mu\rho} \} 
-\{ H_{\rho\sigma} \,,\, B_{\mu\nu} \} \,. \cr } \eqn\nexti$$
We now use \six\ and the Jacobi identity to find
$$\D_\mu H_{\nu\rho\sigma} \pm\hbox{Cycl.}  
= -\, \Bigl\{ \, S \, ,  \{ B_{\mu\nu} \,,\, B_{\rho\sigma} \} 
+ \{ B_{\mu\rho} \,,\, B_{\sigma\nu} \} 
+ \{ B_{\mu\sigma} \,,\, B_{\nu\rho} \}\, \Bigr\} \,. \eqn\nextii$$
It now follows from \erf\ that
$$\{ S \,, H_{\mu\nu\rho\sigma} \, \} = 0 \,.  \eqn\gerff$$
where the four-index antisymmetric field strength $ H_{\mu\nu\rho\sigma}$
is given by
$$ H_{\mu\nu\rho\sigma}  \equiv \D_\mu B_{\nu\rho\sigma} \pm\hbox{Cycl.}
 +\{ B_{\mu\nu} \,,\, B_{\rho\sigma} \} 
+ \{ B_{\mu\rho} \,,\, B_{\sigma\nu} \} 
+ \{ B_{\mu\sigma} \,,\, B_{\nu\rho} \} \,.\eqn\tger$$  
Equation \gerff\ suggests the existence of a four index hamiltonian
function $B_{\mu\nu\rho\sigma}$ such that 
$$H_{\mu\nu\rho\sigma} = \{ S \,, B_{\mu\nu\rho\sigma} \, \}\,.  \eqn\erff$$
The hamiltonian function $B_{\mu\nu\rho\sigma}$ is expected to be 
defined by moduli spaces of surfaces with four special punctures. 

The general structure of the set of equations is now apparent. We have
a family of hamiltonians $B_\mu, B_{\mu\nu} , \cdots $ that
can be thought of as gauge fields, and a family
of field strengths $H_{\mu\nu} , H_{\mu\nu\rho} \cdots$
built out of the gauge fields.  The typical field strength
$H_{\mu_1\cdots \mu_n}$ involves a covariant derivative $D$ acting on the
gauge field with one less index, plus all possible antibrackets of
gauge fields.  All field strengths are $S$-closed: 
$ \{ S , H_{\mu_1\cdots \mu_n}\} = 0$,  and turn out to be
$S$-exact; $H_{\mu_1\cdots \mu_n}= \{ S , B_{\mu_1\cdots \mu_n}\}$.

\chapter{Generalized BRST operator}

A string field theory formulated around an arbitrary background must
have an action linear in the fluctuating string field and 
would be expected to look like
$$S(\Psi )=  S_0 + \langle \Psi,
\F  \rangle + \half \langle \Psi , {\cal Q} \Psi \rangle + 
{1\over 3!} \langle \Psi , [\Psi , \Psi ] \rangle \cdots . \eqn\act$$
This action is expected to be invariant under  gauge 
transformations of the form
$${\delta}_\Lambda \Psi = {\cal Q} \Lambda + [\Psi , \Lambda ]
+ \cdots ,\eqn\gtrn$$
In the above equations ${\cal F}$ is a ghost number three, grassmann odd,
string field, and ${\cal Q}$ is a grassmann odd, ghost number one operator.
The first few conditions for gauge invariance of the above string action read
(Ref.[\zwiebachlong], sect.4.5)
$$  {\cal Q}\ket{{\cal F}} = 0 , \eqn\clfornzero$$
$${\cal Q}^2 \ket{A} = -  \l {\cal F} , A \r \, ,\eqn\failnilp$$
$$ {\cal Q} \l A_1 , A_2 \r  + \l {\cal Q} A_1 , A_2 \r + (-)^{A_1} \l A_1,
{\cal Q}A_2 \r  = -  \l {\cal F},A_1,A_2 \r  . \eqn\foxxfxo$$
When ${\cal F}=0$ we recover the structure arising from conformal backgrounds
and ${\cal Q}$ becomes the BRST operator of the conformal field theory. 
To first approximation (in the departure from conformality)
one expects the string product $[~ ,~ ]$ to be 
defined by the symmetric three punctured sphere that
gives rise to the three string vertex. Similarly the product $[~,~,~]$ should
be defined, to first approximation, by the collection of four punctured
spheres that comprises the four string vertex.

The first of the above identities says that
${\cal Q}$ annihilates the special state $\ket{{\cal F}}$. This is
the analog of the Bianchi identity in Einstein's gravity, as explained
in the introduction. 
The second identity says that as an operator ${\cal Q}^2$ is an operator
roughly represented by the two punctured sphere obtained by
filling one of the punctures of the three string 
vertex with the state $\ket{{\cal F}}$. It is hard to see how the square
of an operator defined by a conventional contour integral 
could be represented by the action of a symmetric three punctured sphere. 
The last
equation is also peculiar in that the failure of ${\cal Q}$ to act like a
derivation of the string product defined (roughly) by the three-string vertex 
is given by the insertion of $F$ on roughly the four-string vertex.
A direct construction of these objects using 2-dimensional field theory
seems difficult. 

In this section we will use the above discussion to 
motivate a set of identities
that could be expected to arise in a two-dimensional field theory.
These identities will involve an operator $Q$, to be referred to as
the generalized BRST operator, and a special string field denoted as $F$.
The consistency of the postulated identities will be examined. Finally
we will examine BRST action on moduli spaces of surfaces with ordinary
and special punctures. The explicit construction of $Q$ and $F$ using
two-dimensional field theory will be discussed in [\zwiebachweyl].

\section{$(Q , F)$ - Descent Equations}

If we are to use two-dimensional field theory ingredients
to build a generalized BRST charge, we expect to build it
by integration over some closed curve $\gamma$
of an operator valued one-form on a two-dimensional surface.
This surface is equipped with a conformal structure and a Weyl factor.
Similarly we can expect $\ket{F}$ to correspond to some ghost number
three local operator $F(z, \bar z)$ of the two-dimensional field theory.

This generalized BRST charge will lose many of the properties
standard in the conformal case, but 
it will do so in a well-defined fashion. A fundamental change is
that the BRST charge will be contour dependent, and we thus write 
$Q(\gamma)$ to indicate
explicitly this dependence. We demand, however, that it should not 
depend on the parametrization of the contour, nor on the local 
coordinates used to do the line integral: the charge arises as the integral
of a well-defined (current) one-form. The failure of contour independence
for homologous contours is equivalent to lack of conservation for the current
one-form whose integral defines the BRST charge.

According to Eqn.\clfornzero\ we should expect  the BRST charge
to annihilate the local operator $F(z,\bar z)$  in some 
suitable sense.
Assume $F$ is inserted at point $P$ using a local coordinate $z$ vanishing
at $P$: $z(p) =0$. Consider then the family of contours 
$\gamma_r = \{ |z| = r \}$ that surround the point $P$ (see Figure 3). 
We demand that
$$\lim_{r\to 0} Q (\gamma_r)\, F (0) =0 \,.  \eqn\vanish$$
This is the best we can do given that the BRST operator is not
conserved.

\Figure{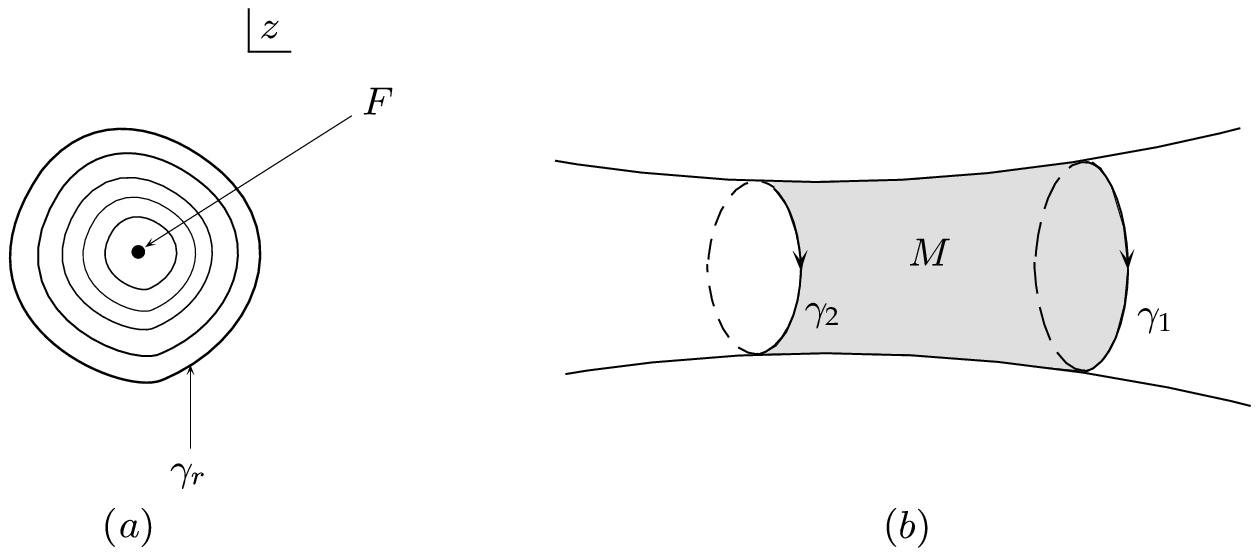}{Figure 3.~(a) The operators $Q(\gamma_r)$, in
the limit $r\to 0$ annihilate the operator $F$ (b) The BRST charge
is not conserved, the difference between the charge evaluated for
two homologous contours $\gamma_1$ and $\gamma_2$ 
is given by a surface integral of $F$ over the region $M$.}

The next identity we postulate has to do with the failure
of $Q$ to square to zero and is thus related to \failnilp. 
This property must again refer
to some chosen contour.
Let $F^{\,[1]}(z,\bar z)$ denote
the operator-valued one-form arising from the local operator $F (z, \bar z)$. 
We now demand
$$ [ Q (\gamma) ]^2 = \, {1\over 2\pi i}\, \int_\gamma F^{\,[1]} \,. \eqn\jkm$$
The integral in the right hand side is over the  contour that defines
the BRST charge in the left hand side. 
Note that ghost number works out, $F^{\,[1]}(z,\bar z)$
is of ghost number two, the same ghost number as
that of $Q^2$. 

Finally, we must consider Eqn.\foxxfxo. This equation addresses
the curve dependence of the BRST charge. As the right hand side 
indicates, the local operator $F(z, \bar z)$ is required to control the
contour dependence. Therefore, 
given two homologous contours $\gamma_1$ and $\gamma_2$,
namely two countours bounding a cylindrical region $M$: $\partial M =\gamma_2
-\gamma_1$ we postulate that
$$Q(\gamma_2) - Q (\gamma_1) =
 -\, {1\over 2\pi i}\, \int_M F^{\,[2]}\,, \eqn\consey$$ 
where $F^{\,[2]}(z, \bar z)$ is the two-form associated to $F$. The ghost
number of  $F^{\,[2]}$ is $+1$, and thus both sides of 
the above equation have the same ghost number. 

The three relevant equations \vanish\jkm, and \conse\ resemble
descent equations, since they involve the various forms that arise
from the local operator $F(z, \bar z)$. We will now rewrite these
equations in a language which is more suitable  for string field theory.
This language will also facilitate the
discussion of consistency checks between the various equations.

\section{$(Q , F)$ Equations in Symplectic Notation}

We will now rewrite equations 
\vanish,\jkm, and \conse\ in the language of antibrackets and
hamiltonian functions.  For this purpose we must introduce
a symplectic structure  in the
space of local operators of the two dimensional field theory. 
This is defined, in analogy to the conformal case, by a correlator
on a special two punctured sphere.

\noindent
\underbar{Canonical Cylinder}. This is the canonical two punctured
Riemann sphere equipped with a special Weyl metric. We recall that
the canonical two punctured sphere is the sphere with  local coordinates
$z_1$ and $z_2$, satisfying $z_1 z_2 =1$, and with the punctures at
$z_1=0$ and $z_2=0$. The Weyl metric $\rho$, with 
$ds= \rho |dz|$, is given by $\rho(z_1) = 1/|z_1|$ (and $\rho(z_2) = 1/|z_2|$).
This metric makes the surface into a
flat infinite cylinder of circumference $2\pi$.

\noindent
\underbar{Symplectic Structure} Acting on two local operators of the
theory it furnishes a number. This number is  the correlator on the canonical
cylinder of the local operators,  inserted
at the punctures, and a line integral of the ghost field  producing
the $c_0^-$ insertion familiar in the conformal case. 

\noindent
\underbar{BRST operator}.
We must now define a BRST operator and the associated BRST hamiltonian function. 
We will denote both the BRST operator
and the BRST hamiltonian
by the same symbol $Q$; which one is being used should
be clear by the context. Given that we have contour dependent
BRST operators $Q(\gamma)$, defining a fixed BRST operator entails
a choice of contour in some chosen surface.   We define the BRST operator
$Q$ to be the operator $Q(\gamma)$ that arises when we choose 
$\gamma$ to be the closed geodesic $|z_1|=|z_2| =1$ in the canonical
cylinder. This will be 
{\it the} BRST operator to be used in string field theory.

We will use the notation of bras and kets. The symplectic form
will be denoted as $\bra{\omega_{12}}$, with $\bra{\omega_{12}}= 
-\bra{\omega_{12}}$, and its inverse as $\ket{S_{12}}$ with $\ket{S_{12}}
= \ket{S_{21}}$. The hamiltonian function $Q$ is given as
$Q = \half \bra{\omega_{12}} Q^{(2)} \ket{\Psi}_1 \ket{\Psi}_2$, where
the $Q$ appearing in the right is the BRST operator. All these results
are familiar in the conformal case. Most results involving
antibrackets will carry from the conformal case to the present
case, one must only make sure
the Weyl metrics of all surfaces are well defined.

We now claim that the following three equations encode in symplectic
language the content of Eqns.\vanish,\jkm, and \consey\  
$$\{ Q\, , B_F^{(2)} \} = - f ( \T^2_1 ) \,, \eqn\th$$ 
$$\half \{ Q\, , Q \} = f (\V'_3) \,,  \eqn\qg$$ 
$$\{ Q\, ,f(\Sigma)\}= f (\K \Sigma)\,.\eqn\qss$$

Let us begin from the last equation. In the conformal case the
right hand side is equal to zero, as a result of the identity
$\bra{\Sigma} \sum Q =0$ which  is proven by contour deformation. 
This time $\Sigma$ is assumed to 
be equipped with a Weyl metric such that the coordinate disks are semiinfinite
canonical cylinders. This is necessary for sewing compatibility since the BRST
operator is defined on the canonical cylinder. The right hand side of
\qss\ is intuitive, it simply indicates that we must integrate
the insertion of $F$ over the two-dimensional region outside of the
coordinate disks. This equation follows from
\consey\ rather simply. Consider the definition of the function $f(\Sigma)$
$$f(\Sigma) = {1\over n!} (-2\pi i)^{3-n}\,\bra{\Sigma}\Psi\rangle_1\cdots \ket
{\Psi}_n \,, \eqn\surfstate$$
A straightforward calculation gives
$$\{ Q \,, \, f(\Sigma)\} = -{1\over n!} (-2\pi i)^{3-n}\, 
\bra{\Sigma}\bigl( \sum Q \bigr)\ket{\Psi}_1\cdots \ket{\Psi}_n \,.
 \eqn\almt$$ 
On the other hand, \consey\ implies that $\sum Q = {1\over 2\pi i}
\int_{\Sigma- \cup D_i} F^{[2]}$ (recall the contours for $Q$ are
oriented oppositely to the boundary of the surface minus the unit disks).
We thus find 
$$\eqalign{
\{ Q \,, \, f(\Sigma)\} &= {1\over n!} (-2\pi i)^{3-(n+1)}\hskip-9pt 
\int_{\Sigma - \cup D_i}\bra{\Sigma ; p }F^{[2]}\ket{\Psi}_1
\cdots \ket{\Psi}_n \cr
&= {1\over n!}\hskip-9pt 
\int_{\Sigma - \cup D_i}\bra{\Omega^{[2]n+1} }F\rangle_{n+1}\ket{\Psi}_1
\cdots \ket{\Psi}_n = f(\K\Sigma)\,.  \cr }\eqn\almt$$ 

Let us now consider Eqn.\qg. It expresses the failure of the BRST
hamiltonian to have zero bracket with itself in terms of the insertion of
a single $F$ in the special puncture of the three punctured sphere $\V'_3$.
This three punctured sphere is taken to be the canonical cylinder with
the third punture considered special and located at $z_1=1$. 
The coordinate at that puncture
must be chosen such that the cylinder is symmetric under the exchange of
the two ordinary punctures.  To verify Eqn.\qg\ we begin with the
computation of the left hand side. One finds
$$\half \{ Q , Q \} = \half \bra{\omega_{12}}  (Q Q \ket{\Psi})_1 \ket{\Psi}_2
 = \half \,\bra{\omega_{12}} 
 \Bigl(\,{1\over 2\pi i}\oint F^{[1]} \ket{\Psi}\Bigr)_1 \ket{\Psi}_2\,,
\eqn\hlk$$
where use was made of \jkm. We now claim that
$$\Bigl(_{\hskip -13pt 1'}\,\,\,\oint F^{[1]} \Bigr)_1  = 
 2\pi i\,\bra{\V'_{123}} F\rangle_3 \ket{S_{1'2}}\,, \eqn\fclaim$$ 
where the left hand side refers to the line integral over the 
central geodesic of the canonical cylinder using the local coordinates
induced by the special puncture of $\V'_3$. The right hand side creates
this integration by inserting the state $F$ in the special puncture
and then by rotating it via the twist-sewing ket $\ket{S}$. In \fclaim\
the state $F$ is arbitrary. The only
part of \fclaim\ that needs verification is the constant of proportionality.
The right hand side can be compared to the left hand side by separating
out the $b_0^{-(2)}$ factor in the twist sewing ket and expressing
$\bra{\V'_{123}}b_0^{-(2)}$ in terms of antighost operators in the
state space of $\ket{F}$. This small calculation confims the
value of the proportionality constant in \fclaim.
We can now use \fclaim\ in \hlk\ and find
$$\eqalign{
\half \{ Q , Q \} 
 &= \half \,\bra{\omega_{1'2}} 
 \,\bra{\V'_{12'3}} F\rangle_3 \ket{S_{1'2'}} \ket{\Psi}_1 \ket{\Psi}_2\,, \cr
&= \half \,
\bra{\V'_{123}} F\rangle_3 \ket{\Psi}_1 \ket{\Psi}_2\, \cr
& = f(\V'_3)\,, \cr} \eqn\huk$$
as we wanted to show.  We now consider equation \th, which as we will see
is a consequence of \vanish\ and \consey. A short calculation shows that
$$\{ Q , B_F^{(2)} \} = \bra{\omega_{12}} \Psi\rangle_1 \, \bigl( Q \ket{F}
\bigr)_2 = - {1\over 2\pi i} \,\,\bra{\omega_{12}} \Psi\rangle_1 \, 
\Bigl(\,\,\int_D F^{[2]} \, \ket{F} \Bigr)_2 \,, \eqn\ned$$ 
where $D$ denotes the punctured disk $0<|z_2| \leq 1$
corresponding to the semiinfinite portion
of the canonical cylinder that contains puncture number two, where $\ket{F}$ is
inserted. In obtaining this we have used the contour deformation property
\consey\ to relate the BRST operator, defined on the central geodesic, to
the operator $Q(\gamma_r)$ where $\gamma_r$ is the geodesic  $|z_2| =r$
in the limit as $r\to 0$ (see Figure 4). 
In this limit $Q(\gamma_r)$ annihilates $F$ 
by virtue of \vanish\ and
we simply get the surface integral of the $F$ insertion. 

\Figure{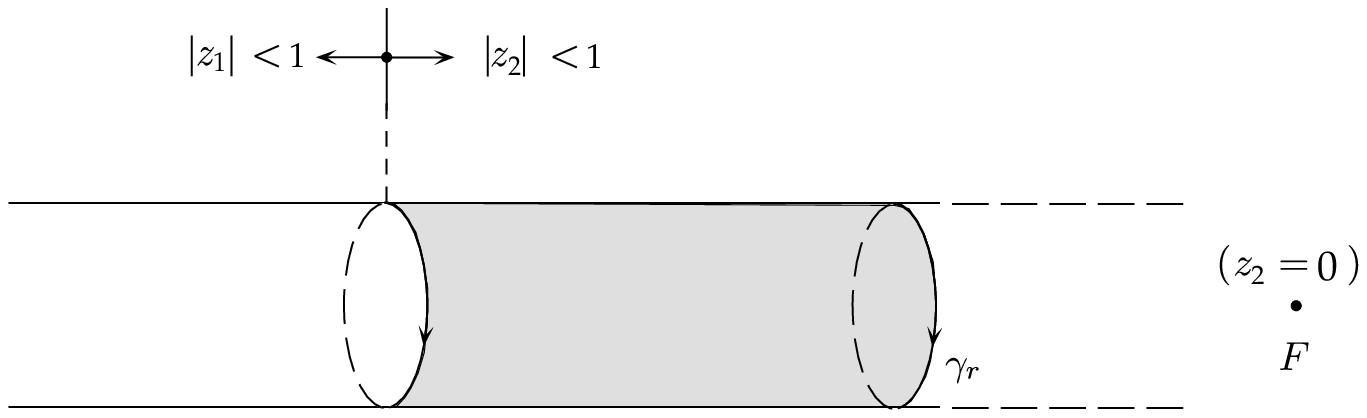}{Figure 4.~In calculating $\{Q, B_F^{(2)} \}$
we find the BRST insertion on the central geodesic. This operator
can be moved towards $z_2=0$, where $F$ is inserted, and in
the process we obtain the surface integral of $F$ over the shaded region.}

In order to 
proceed further we now claim that for arbitrary states $\ket{F}$ and $\ket{\O}$
the following equation holds
$$\Bigl(\,\int_D F^{[2]} \,\ket{\O}\,\Bigr)_{1'}  = 
\, 2\pi i \int_{\tau^2_1 (0)} 
\bra{\,\Omega^{[1]}_{1\bar 1\bar 2}\,} F\rangle_{\bar 1}
 \ket{S_{11'}} \ket{\O}_{\bar 2} \,, \eqn\sclaim$$ 
where $\tau^2_1 (u)$ is the one (real) dimension space of three punctured
spheres introduced in [\zwiebachhms] and reviewed in section two. Roughly
speaking the integration over $\tau^2_1(0)$ generates the radial part of the
integration over the disk, while the twist-sewing ket $\ket{S}$ generates
the angular part of the integration. Equation \sclaim\ can be verified 
by explicit construction of both the left hand side and the right hand
side (for the right hand side, part of the construction can be found
following Eqn.(4.24) of Ref.[\zwiebachhms]). Back in \ned\ we now obtain
$$\eqalign{
\{ Q , B_F^{(2)} \} &=  - \,\bra{\omega_{12}} \Psi\rangle_1\hskip-6pt
 \int_{\tau^2_1 (0)}\hskip-6pt 
\bra{\,\Omega^{[1]}_{0\bar 1\bar 2}\,} F\rangle_{\bar 1}
 \ket{S_{02}} \ket{F}_{\bar 2} 
 \, =\, - \hskip-6pt\int_{\tau^2_1 (0)}\hskip-6pt 
\bra{\,\Omega^{[1]}_{1\bar 1\bar 2}\,}\Psi\rangle_1 \ket{F}_{\bar 1}
 \ket{F}_{\bar 2}\cr
& = -\,\half\, \int_{\T^2_1} 
\bra{\,\Omega^{[1]}_{1\bar 1\bar 2}\,}\Psi\rangle_1 \ket{F}_{\bar 1}
 \ket{F}_{\bar 2} = - f(\T^2_1)\,. \cr } \eqn\nedd$$ 
This concludes our discussion of the basic BRST/Weyl identities
in the antibracket formalism.

\section{$Q$-Action on moduli spaces of surfaces}

We must now consider the generalization of the property  
$\{ Q\, ,f(\Sigma)\}= f (\K \Sigma)$, describing BRST action on the correlators
on a fixed surface,
to the case when we deal with  correlators integrated over moduli spaces 
of surfaces. We will consider first the case of moduli spaces
of surfaces with ordinary punctures only. We then turn to the
case of moduli spaces of surfaces with both ordinary and 
special punctures.

\noindent
\underbar{Moduli spaces with ordinary punctures}
Let $\A$ be a a moduli space of surfaces with just ordinary punctures.
Recall that surfaces now means Riemann surfaces with punctures, local
coordinates at the punctures, and a Weyl factor. We will always assume
that the coordinate disks around ordinary punctures are semiinfinite
flat cylinders of circumference $2\pi$.  We now demand that
$$\eqalign{
 \{ Q , f(\A ) \} &= - f ( \p \A ) + (-)^\A f( \K \A ) \,  \cr 
& =  - f \Bigl( \p \A  - (-)^\A  \K \A \Bigr) \,. \cr } \eqn\sp$$
This equation cannot be fully established
starting from the BRST/Weyl equations
we have postulated so far. It must be derived from an explicit construction
of $Q$ and the explicit construction of forms in the moduli space of
punctured Riemann surfaces equipped with Weyl metrics. These topics
will be discussed in [\zwiebachweyl].
In the conformal limit $F=0$  the second term in the right hand side
vanishes, and we 
recover the familiar correspondance $Q \leftrightarrow -\p$ that
establishes that BRST acts as an exterior derivative on CFT-valued
forms on moduli spaces of Riemann surfaces. The second term in the
right hand side is the natural extension of the term appearing in
\qss. Its sign factor arises from the sign factor in Eqn.(2.10) of
Ref.[\zwiebachhms]. 

The $Q/\partial$ correspondance does not hold in the
non-conformal case. Still,  $Q$ has a moduli space interpretation
as the operator $-\partial \pm \K$, where $\K$ inserts the special state
$F$.  Note that the $\K$ insertion of $F$ roughly amounts to a 
covariant derivative $D_F (\Gamma)$ along the direction $F$ with
the special connection $\Gamma$. Thus we have the
correspondance 
$Q -   D_F \,\leftrightarrow\, -\p$.

It is simple to extend \sp\ to the case when one of the ordinary punctures
has a special state $\O$ inserted in it (rather than the string field).
The operator 
$B_\O^{(2)}$ defined in \lham\ does the insertion. 
The Jacobi identity applied to
$\{ Q, \{ f(\A), B^{(2)}_\O \} \}$ yields the equation
$$ \{ Q , f_\O (\I \A ) \} =\{\,  \{ Q,  f(\A) \} , B^{(2)}_\O \, \}
 + (-)^\A f_{Q\O} (\I\A )\,, \eqn\qyp$$
and as a result we obtain
$$ \{ Q , f_\O (\I\A ) \} = - f_\O ( \I\p \A )  + (-)^\A f_{Q\O} (\I\A ) 
+ (-)^\A f_{F\O} (\I \K \A ) \,,\eqn\qsp$$
where the $\K$ inserts the $F$ state. Here $f_{Q\O}(\I\A)$ represents
the correlator where $\O$ is inserted in the special puncture created
by $\I$, and there is a BRST insertion on the boundary of the
coordinate disk.

\noindent
\underbar{Moduli spaces of surfaces 
with special punctures}
Consider a general  $\B$-type space having a number of special punctures.
The case of interest in the present paper is that where
the state $F$ is inserted in all of the special punctures. 
The Weyl metric on the coordinate
disk of a special puncture taken to be arbitrary. 
In analogy to \sp\ we now find 
$$ \{ Q , f (\B ) \} = - f ( \p \B ) 
+ (-)^\B f_{F\cdots} (\K \B )\,. \eqn\qssp$$
This equation requires explanation. Recall~that $Q$ makes its way into
$f(\B)$ through the ordinary punctures, the antibracket 
introducing the sum of correlators each of which has a $Q(\partial D_i)$, with
$D_i$ the disk associated to an ordinary puncture.
Imagine now introducing a small contour $\gamma_\epsilon$ around each
of the special punctures, with $\epsilon$ chosen small enough that
all such contours are disjoint. Now add and subtract to the left hand
side of \qssp\ correlators where we have
a  $Q(\gamma_\epsilon)$ acting on each of the special punctures.
As we now do contour deformation through the surface to cancel out all
the $Q$'s we obtain the insertion of $F$ over the surface minus the
coordinate disks $D_i$ and minus the $\epsilon$-disks
around the special punctures. In addition to this, since we added and
subtracted terms,  we get a sum of correlators
each with a $Q(\gamma_\epsilon)$ around a special puncture.  We now
take the limit as $\epsilon\to 0$ and the latter terms vanish because
the BRST integrals are closing into $F$ states (recall \vanish).
The former term gives the $F$ insertion over the surface minus the
unit disks around the ordinary punctures. This is precisely the definition
of the operator $\K$ on moduli spaces with special punctures; insertion
ignoring the presence of the special punctures. This 
explains Eqn.\qssp\ except for the sign factor and the dots. By the dots
we mean that the $F$ state to be inserted by the $\K$ insertion appears
to the left of all other $F$ states. This is the reason the sign factor
has not changed relative to \sp. Since in our convention 
$\K$ uses the last label and the $F$ states are arranged in increasing
label value from left to right, we must move the last $F$ state across
the other $F$'s. Letting $k$ denote the number of special punctures, the 
sign factor in \qssp\ becomes
$$ \{ Q , f(\B ) \} = - f \Bigl( \p \B  -(-)^{k+\B}\K \B \Bigr) \,,\eqn\qaww$$
and this holds for arbitrary $\B$ spaces. If the $\B$ space is a string vertex,
then $k=\hbox{dim}~\B$(mod $2$) and we find
$$ \{ Q , f(\B ) \} = - f \Bigl( \p \B  -\K \B \Bigr) \,.\eqn\qaw$$

\section{Consistency checks}

It is useful to perform a few consistency checks with the equations
we have introduced in the last two subsections. These checks basically
use the Jacobi identity \jacobin\ for the antibracket, and the 
properties of the operators $\K$ and $\I$.
 
The state $\ket{F}$ being of ghost number three cannot couple to itself 
through the antibracket. Thus $\{ B_F^{(2)}, B_F^{(2)} \} =0$.  
This result is compatible with \th\  and  the Jacobi identity:
$$0= \Bigl\{ \{ B_F^{(2)} , B_F^{(2)} \}\, , \, Q\Bigr\} \sim 
\Bigl\{ \{ Q , B_F^{(2)} \}\, , \, B_F^{(2)} \Bigr\} \sim \{ f(\T^2_1)
\, , B_F^{(2)} \} \sim f(\I\T^2_1) =0\,, \eqn\gjd$$
where the final equality follows because of ghost number: 
the sphere needs seven units of 
ghost number, the three $F$'s provide nine.

Equation \qg\ is tested by evaluating $\{ \half \{ Q, Q\} , B_F^{(2)} \}$ in
two different ways.  Direct evaluation gives 
$$\Bigl\{ \, \half \{ Q, Q\}\, ,\, B_F^{(2)} \Bigr\} =  \{ f (\V'_3)\,,
B_F^{(2)}\, \}
=  f (\I\V'_3) \,,\eqn\rte$$
on the other hand, using the Jacobi identity 
$$\eqalign{
\Bigl\{ \, \half\{ Q, Q\} , B_F^{(2)} \Bigr\} &=  
\,\,\Bigl\{ \, Q\,, \,\{ Q , B_F^{(2)}\}  \Bigr \}\,, \cr
& = -\,  \{ Q \, , \, f(\T^2_1)\} \,, \cr
&=  f(\p \T^2_1 + \K\T^2_1) \,,\cr }\eqn\ert$$
where used was made of \qaww\ in the last step. 
Using \boundtau\ we see that the first term in the right hand
side gives the desired answer. On the other hand we expect $\K\T^2_1=0$.
As we will explain precisely in sect.5.2, 
one can think of $\T^2_1$ as 
$\K \B^1_1$, where $\B^1_1$ is a canonical cylinder with one special and
one ordinary puncture. The vanishing of $\K\T^2_1$ then follows 
from $\K^2=0$.  This implies consistency.

As a final, and more general consistency check we now evaluate 
 $\bigl\{ \half\{ Q, Q \} ,f(\B) \, \bigr\}$ in two different ways.
By direct evaluation we find
$$\Bigl\{ \half\{ Q, Q \} ,f(\B) \, \Bigr\} = \{ f(\V'_3) , f(\B)\} = 
- f\Bigl( \{ \V'_3 , \B \} \Bigr)\, . \eqn\fway$$
On the other hand, using the Jacobi identity and recalling that $f(\B)$ is
Grassmann even we get: 
$$\eqalign{
\Bigl\{ \, \half\{ Q, Q\} , f(\B) \Bigr\} &=  
\,\,\Bigl\{ \, Q\,, \,\{ Q , f(\B)\}  \Bigr \}\,, \cr
& =\,-\Bigl\{ \, Q\,, f \bigl( (\partial - \K) \B \bigr) \Bigr \}\,\,, \cr
& =\,  f \, \Bigl( (\partial +\K) (\partial - \K) \B \Bigr) \,, \cr
&=  - f\, \bigl(\, [\, \partial , \K\, ]\,  \B\,  \bigr) \,,\cr
&=  - f\, \bigl(\, \{ \V'_3 , \B\,\}     \bigr) \,,\cr }\eqn\ert$$
in agreement with \fway. In deriving \ert\ we made use of \qaww, of the
properties $\partial^2 = \K^2 =0$ and of the second equation in \knewd.

\goodbreak %
\chapter{\bf Constructing the New String Action: First Few Terms}

We will now take the first few steps in the
construction of the string field action. The construction will be
perturbative in $F$. Moreover, it will make  no reference to any conformal
theory. We will denote by $S_0$ the $F$ independent terms
in the complete action, by
$S_1$ the action including $F$ independent terms and terms linear in $F$, by 
$S_2$  the action
including all terms of order less than or equal to quadratic in $F$, and so on.

\section{The Construction to Zeroth-order}

The discussion in the previous section showed that
the new BRST operator fails to satisfy the identities of the
standard CFT BRST operator, but that the failure is through terms
linear in $F$.
We can therefore begin the construction by taking $S_0$ to have
the same form it had in the conformal case:
$$S_0 =  Q + f(\V) \,.\eqn\szero$$
It should be emphasized that this $Q$ is the one defined for
the chosen two-dimensional non conformal theory, and similarly the
correlators in $f(\V)$ are correlators in the non conformal theory.
$S_0$ has nothing to do with a conformal theory. In fact the chosen
non conformal theory need not be close to any CFT. 

We compute the antibracket of this action with itself
$$\eqalign{
\{ S_0 \,,\, S_0 \,\} &= \{ Q , Q\} 
+ 2\, \{ Q , f(\V) \} + \{ f (\V) , f(\V) \}\,,\cr
&= 2f (\V'_3) - 2 \, f\bigl(  \partial \V - \K\V \bigr) \,
 - f( \{ \V , \V \} )\,, \cr} \eqn\szeroa$$
where use was made of Eqns. \qg\ and \sp. The result is  
neatly organized as
$$\eqalign{
\{ S_0 \,,\, S_0 \,\} &= - 2\, f \Bigl( \p\V + \half \{ \V , \V \}  \Bigr)\cr
&\quad  + 2\, f \Bigl( \V'_3 + \K \V \Bigr)\,, \cr} \eqn\yy$$
where the first line in the right hand side involves moduli
spaces with no special punctures, and the second line involves
moduli spaces with one special puncture.
The geometrical recursion relations \recrel\ for the string vertices $\V$ 
imply that the first line in the above right hand side vanishes. We  find,
as expected,
that $S_0$ is correct to zeroth order in the string field $F$. We 
therefore obtain 
$$\{ S_0 \,,\, S_0 \,\} =  2\, f \Bigl( \V'_3 + \K \V \Bigr) \,. \eqn\xx$$
This nonvanishing right hand side signals that $S_0$ is not fully
consistent.
We must change the action by terms having $F$ insertions. The interpolating
$\B^1$ spaces of background independence are the natural candidates.

\section{The construction to first order}

We have emphasized earlier that the $\B^1$ spaces are natural for the
insertion of the ghost number three state $F$. This state is inserted
on the special puncture, and string fields are inserted on the ordinary
punctures. Correlators will exist when the total ghost number of the
string fields is two times the number of ordinary punctures, just
as it is the case for string vertices. We therefore write 
$$\eqalign{
S_1 &=  S_0 + \Bigl(  - B^{(2)}_F+ f_F(\B^1)\Bigr)  \, , \cr
&=   Q + f(\V) 
+ \Bigl( - B_F^{(2)}+ f_F(\B^1)\Bigr) \,. \cr} \eqn\sone$$
and claim that $S_1$ is the correct action to order $\O(F)$.
To verify this we now calculate $\{ S_1 , S_1\}$
$$\eqalign{
\{ S_1 , S_1\} &= \,\,\{ S_0 ,  S_0 \} \cr
&\quad + 2\,\{ \, S_0\, ,\, -B^{(2)}_F + f(\B^1) \}\cr
&\quad +
\Bigl\{ -B^{(2)}_F + f(\B^1)\,,\, -B^{(2)}_F + f(\B^1) \Bigr\}\,. \cr}
\eqn\sonec$$
We must calculate the various pieces in this equation. The first line
has already been computed. The second line requires the evaluation of
two terms, the first of which is  
$$\{ S_0 , B^{(2)}_F \} = \{ Q , B^{(2)}_F\} + \{f(\V) , B^{(2)}_F \}
= f(\,-\T^2_1 +\I \V\,) \,, \eqn\fpart$$
where use was made of \th\ and \iopin. Note that in the right
hand side the first contribution is
from a space with two special punctures, and the second contribution is
from a space with a single special puncture. The second term we need
in this line is
$$\eqalign{
\{ S_0 , f(\B^1)\} &= \{ Q , f(\B^1) \} + \{ f(\V) , f(\B^1) \} \cr
&= - f (\partial \B^1 - \K\B^1 + \{ \V , \B^1\} ) \,,\cr
&= - f (\delta_\V\B^1 - \K\B^1 )\,. \cr} \eqn\midcomp$$
where use was made of \qaw.  The third line gives
$$\eqalign{
\Bigl\{ -B^{(2)}_F + f(\B^1)\,,\, -B^{(2)}_F + f(\B^1) \Bigr\}& =
- 2\, \{  B^{(2)}_F \,,\, f(\B^1) \}  +  
\{  f(\B^1)\,,\, f(\B^1) \}\,, \cr
&= - 2 f(\I \B^1) - f ( \{ \B^1 , \B^1 \} ) \,, \cr} \eqn\secpart$$
given that the antibracket of $B^{(2)}_F$ with itself is zero,
and making use of  \iopin\ and \gtnss. Putting it all together
$$\eqalign{
\{ S_1,S_1\} & = ~~\, 2\, f\Bigl( \,\V'_3 +(\K -\I )\V -\delta_\V \B^1 \Bigr)\cr
&\quad +2\, f \Bigl( \, \T^2_1 + (\K - \I) \B^1 - \half \{ \B^1 , \B^1 \}
\Bigr) \, . \cr } \eqn\dlsk$$
In the right hand side we show in the first line the terms with one special
puncture. We see that as desired they cancel out by virtue of the 
recursion relations \gmallow\ for the $\B^1$ spaces. Thus we have succeded
in defining an $S_1$ that satisfies the master equation to first nontrivial
order in $F$. It is not a complete answer because we have
$$\{ S_1,S_1\}  =
\, 2\, f \Bigl( \, \T^2_1 + (\K - \I) \B^1 - \half \{ \B^1 , \B^1 \}
\Bigr) \,\not= 0 . \eqn\dlsk$$
It is clear that we must add to $S_1$ a term corresponding
to surfaces with two special punctures. 

\section{Construction to second order}

To continue to second order in $F$ we now simply include the
$\B^2$ spaces discussed in [\zwiebachhms] and reviewed in sect.2.
We now claim that 
$$S_2 = S_1 + f (\B^2)\,, \eqn\nextcorr$$ 
is the correct action to order $\O(F^2)$.
To show this we compute $\{ S_2 , S_2 \}$
$$\eqalign{
\{ S_2 , S_2 \} &= \{ S_1 , S_1 \} + 2 \{ S_1 ,f(\B^2) \}+\O (F^4)\,,\cr
&= \{ S_1 , S_1 \} + 2 \{ S_0 ,f(\B^2) \}+\O (F^3)\,,\cr 
&=\{ S_1 , S_1 \} - 2 f (\delta_\V\B^2) +\O' (F^3) \cr } \eqn\bhg$$
since the antibracket of $f(\B^2)$ with itself has four special punctures
and the antibracket of $f(\B^2)$ with $f(\B^1)$, or $B^{(2)}_F$ has three
special punctures.  We therefore
have
$$
\{ S_2,S_2\}  =
\, 2\, f \Bigl( \, \T^2_1 + (\K -\I) \B^1 - \half \{ \B^1 , \B^1 \}
- \delta_\V \B^2 \, \Bigr) \, + \O (F^3) . \eqn\dlskk$$
The recursion relation \gmallow\ for $\B^2$ spaces implies that
the part of $\{S_2 , S_2 \}$ quadratic
in $F$ vanishes. Thus $S_2$ is accurate to order $F^2$.  

\chapter{Constructing the New String Action: All orders}

In order to complete the construction of a gauge
invariant action we need moduli spaces with more than
two special punctures. We have indicated in sect.2.3 how such
spaces would arise from higher order background independence
consistency conditions. It is simpler, however, to derive 
the equations that such spaces must satisfy directly from 
gauge invariance, and,  then show that the spaces can be
consistently defined. This is how we will proceed. We will
see how string vertices and $\B$ spaces can be naturally 
grouped together into a single ``space" satisfying  simple
recursion relations.   Finally, we will introduce the
moduli space $\B^1_1$, and show how it simplifies the 
writing of recursion relations. 

\section{Construction to all orders}
Consider now  general $\B^k$ moduli spaces of surfaces with $k$ 
special punctures.
As before  we will define $\B^{k}= \sum_n \B^k_n\,$
where each of the $\B^k_n$ spaces has $k$ special labelled punctures
 $(\bar 1, \bar 2, \cdots \bar k)$, and $n$ ordinary punctures. 
When $k=0$ the  moduli spaces in
$\B^0$ are identified with the string vertices, and in this case  $n\geq 3$. 
When $k=1$ we have
the moduli spaces that appeared in the analysis of background independence
and here $n\geq 2$ (this will change in the present analysis). 
Our studies in Ref.[\zwiebachhms] show that
for $k=2$ we have $n\geq 1$. The real
dimensionality of $\B^k_n$ exceeds by $k$ that of the moduli space $\M_{k+n}$
of punctured spheres.  
It follows that all moduli spaces in $\B^k$ are odd if $k$ is
odd, and are all even if $k$ is even.

We now claim that the complete gauge invariant action is given as
$$S= S_0 - B_F^{(2)} + \sum_{k=1}^\infty f (\B^k) \,. \eqn\sdes$$
To verify this claim we must check the master equation,
and we must therefore compute
$$\eqalign{
\{ S , S \} &= \{ S_0 , S_0\} - 2 \{ S_0 , B_F^{(2)} \}  \cr
&\quad +  2 \sum_{k=1}^\infty \{ S_0 , f(\B^k ) \} 
-  2 \sum_{k=1}^\infty \{  f(\B^k )\,,B^{(2)}_F  \}\cr
&\quad + \sum_{k_1,k_2 = 1}^\infty \{ f(\B^{k_1} \,,\, f(\B^{k_2}) \}\,.\cr}
\eqn\computea$$
We now use \xx, \fpart, \iopin\ and \gtnss\ to rewrite the above equation as
$$\eqalign{
\bigl\{\, S\, , \,S\, \bigr\} & = 
+2f \Bigl( \V'_3 + \K\V + \T^2_1 - \I\V   \Bigr)  \cr
&\quad + 2 \sum_{k = 1} f\Bigl( -\delta_\V\, \B^k + \,(\K-\I) \B^k\Bigr) \cr
& \quad - \sum_{k_1,k_2 = 1}
f \Bigl( \{ \B^{k_1} \,,\, \B^{k_2} \, \}\Bigr) \,. \cr }  \eqn\ufg$$
It is useful to separate out terms with one and two special punctures,
since such terms are not completely generic. Doing so and rearranging somewhat
the above equation we find
$$\eqalign{
\bigl\{\, S\, , \,S\, \bigr\} & = 
-2f \Bigl( \delta_\V  \B^1 - (\V'_3 + \M\V)  \Bigr)  \cr
&\quad  -2f \Bigl( \delta_\V  \B^2 - ( \T^2_1 + \M\B^1)  \Bigr)  \cr
&\quad - 2 \sum_{k=3}^\infty f\Bigl( \delta_\V\,\B^k\Bigr)    
- 2 \sum_{k = 2}^\infty f\Bigl( -\,\M \B^k\Bigr) \cr
& \quad - \sum_{k= 1}^\infty \sum_{l=1}^k
f \Bigl( \{ \B^{l} \,,\, \B^{k+1-l} \, \}\Bigr) \,, \cr }  \eqn\uffg$$
and with a final rearrangement we write
$$\eqalign{
\bigl\{\, S\, , \,S\, \bigr\} & = 
-2f \Bigl( \delta_\V  \B^1 - (\V'_3 + \M\V)  \Bigr)  \cr
&\quad  -2f \Bigl( \delta_\V  \B^2 - ( \T^2_1 + \M\B^1 - \half \{ \B^1\,,\,
\B^1 \} )  \Bigr)  \cr
&\quad - 2f \Bigl( \,\sum_{k=2}^\infty \Bigl[ \delta_\V\,\B^{k+1}    
- \M \B^k
 +\half  \sum_{l=1}^k
 \{ \B^{l} \,,\, \B^{k+1-l} \, \} \Bigr] \Bigr) \,.\cr }  \eqn\uffg$$
We recognize the familiar recursion relations for $\B^1$ and $\B^2$ spaces,
and find the required recursion relations for $\B^k$ spaces with $k\geq 3$:
$$\eqalign{
\delta_\V\B^1 &=  \V'_3 + \M\V \,, \cr
 \delta_\V  \B^2 &=  \T^2_1 + \M\B^1 - \half \{ \B^1\,,\,\B^1 \} \,, \cr
\delta_\V \B^{k+1}  &= \,
 \M \B^{k} - \half \sum_{l = 1}^k 
\, \bigl\{ \B^{l} \,,\, \B^{k+1-l} \, \bigr\}
 \,. \cr } \eqn\nvss$$
The above identities are better appreciated if we append the recursion relations
for string vertices, and we separate out the $\{ \V , \cdot \, \}$ term
in each $\delta_\V$. We find
$$\eqalign{
\partial\V &= - \half \{ \V \,,\, \V \} \,, \cr
\partial\B^1 &=  \V'_3 + \M\V  - \{ \V , \B^1 \} \,, \cr
\partial\B^2 &=  \T^2_1 + \M\B^1 - \half \{ \B^1\,,\,\B^1 \}
 - \{ \V , \B^2 \}  \,, \cr
\partial\B^{k+1}  &= \,
 \M \B^{k} - \half \sum_{l = 1}^k 
\, \bigl\{ \B^{l} \,,\, \B^{k+1-l} \, \bigr\}
 - \{ \V , \B^{k+1} \}  \,. \cr } \eqn\nvff$$
We now see that it is natural to think of the string vertices $\V$ as 
a the lowest member of the family of $\B$ spaces, a $\B$ space
with zero number of special punctures
$$\B^0 \equiv \V \,.\eqn\spbb$$
We now introduce the complete formal sum of $\B^k$ spaces
$$\B \equiv \sum_{k=0}^\infty \B^k = \V + \B^1 + \B^2 + \cdots\, . \eqn\nwb$$
The whole set of recursion relation \nvff\ is then summarized as as
a single equation for the total $\B$ space
$$\partial \B = \V'_3 + \T^2_1 + \M \B - \half  \{ \B , \B \} \,.\eqn\mnbv$$      
This is a fairly simple equation, simple enough to allow the main 
consistency check. This is the check that $\partial(\partial\B) =0$, or
that the right hand side of the above equation has zero boundary. 
If this consistency check works out one should be able to define the
$\B$ spaces recursively, as was done in [\senzwiebachtwo,\zwiebachhms].

Before starting this verification it is useful to remark
that previous equations derived for single $\B^k$ spaces now hold for the
sum $\B$.  For example,
the Jacobi identity \newjac\ implies that
$$\{ \{ \B , \B \} \, , \, \B \} = 0\, . \eqn\trto$$
Moreover, \mopiden\ implies that
$$[\partial , \M ]\, \B = \{ \V'_3\, , \,\B \} \,. \eqn\mopi$$
Let us now verify  the consistency condition. Since $\V'_3$ has no
boundary we must verify that 
$$ \partial  \T^2_1 + [ \partial , \M ] \B + 
\M (\partial \B) - \{ \partial \B \,,\, \B \} = 0 \,. \eqn\nnbb$$ 
The left  hand side equals:
$$\eqalign{
 & \qquad\, +\I\V'_3 + \{ \V'_3 , \B \} \cr
&\qquad +\M \Bigl(\V'_3 +  \T^2_1 + \M\B  
 - \half \{ \B \,,\, \B \}\Bigr)   \cr
&\qquad -  \Bigl\{ \V'_3 + \T^2_1 + \M \B - \half  \{ \B , \B \}
\,,\, \B   \Bigr\} \cr } \eqn\nmbb$$
Since $\K\V'_3 =0$ (the  unit disks around the
ordinary punctures cover the sphere)  we have $\M\V'_3 = -\I \V'_3$, and this
cancels the first term. Moreover $\M\T^2_1 = \K\T^2_1 - \I\T^2_1 =0$. The
vanishing of $\K\T^2_1$ follows from the fact that $\T^2_1=\K\B^1_1$
(to be explained below), and the vanishing of $\I\T^2_1$ follows from ghost
number, as explained after \gjd. Using the Jacobi identity, we now find
$$\eqalign{
& \qquad + \{ \V'_3 , \B \} \cr
&\qquad -\{ \B , \T^2_1 \}  
 + \{\M \B \,,\, \B \}   \cr
&\qquad -  \Bigl\{ \V'_3 + \T^2_1 + \M \B
\,,\, \B   \Bigr\} \cr } \eqn\nmbb$$
and since the space $\T^2_1$ is effectively odd (dimension one and two
special punctures) $\{ \B , \T^2_1 \} = - \{ \T^2_1 , \B \}$, and
we see that all terms cancel out. This completes our verification
of the consistency conditions. 

\section{Introducing $\B^1_1$}

Equation \sdes\ for the string action would simply read $S= Q+ f(\B)$ were
it not for the $B_F^{(2)}$ term. This term is linear in $F$ and would be
natural to try to include it as part of the $\B^1$ spaces. Since 
$B_F^{(2)}$ is also linear in the string field it must correspond 
to a $\B$-space with one special puncture and one ordinary puncture.
We will denote such space as $\B^1_1$. The reason this space was not 
introduced earlier is that being a two punctured sphere it is somewhat
peculiar. The function associated to $\B^1_1$  is not just the corresponding
surface state with an insertion of a string field and an $F$ state, one
must include a ghost insertion. We simply define
$$f(\B^1_1) \equiv - B_F^{(2)} = \bra{\omega_{12}}\Psi\rangle_1 \ket{F}_2 \,,
\eqn\newb$$
where the last equality follows from \lham.
It is consistent to declare that for an arbitrary $\B$ space
$$\{\B , \B^1_1 \}  =  \I\B\, , \eqn\decl$$
since  \iopin\ and \gtnss\ imply that
$f(\I\B) = \{ f(\B), B_F^{(2)}\} = -\{ f(\B), f(\B^1_1) \} = 
f( \{ \B, \B^1_1 \})$. Since $\I$ does not change the dimensionality
of a space, 
and the antibracket adds one unit to the dimensionality, it follows
from \decl\ that $\B^1_1$ should be thought
as a moduli space of  dimension minus one. Since the surfaces in
the moduli space have one special puncture, the space is effectively
Grassmann even, as it should be if it is to be thought as a string 
vertex.

The picture of the $\B^1_1$ space as a surface is clear, it is the canonical 
cylinder with one puncture declared special (in some sense something
as $\I\V_2$, if we had defined a space $\V_2$ with two punctures).
What happens if $\K$ acts on $\B^1_1$ ?  By our geometrical definition
of $\K$ we would expect it to add a new special puncture throughout the
unit disk of the original special puncture. Due to the conformal isometry
of the sphere this two-dimensional insertion region can be thought simply as
an insertion over a (one-dimensional) line starting from the central geodesic
and going all the way towards the special puncture. But this is nothing
else that the space $\T^2_1$ discussed earlier. We therefore claim that
$$\K\B^1_1 = \T^2_1 \,.\eqn\nnont$$
Note that given that $\T^2_1$ is of dimension one, 
and $\K$ adds two
units of dimension, we confirm that $\B^1_1$ must be thought of 
dimension minus one.  Let us now confirm \nnont\ by more explicit means.
We recall Eqn.\sclaim\ which taking $\ket{\O} = \ket{F}$ reads
$$\Bigl(\,\int_D F^{[2]} \,\ket{F}\,\Bigr)_{1'}  = 
\, 2\pi i \int_{\T^2_1} 
\bra{\,\Omega^{[1]}_{1\bar 1\bar 2}\,} F\rangle_{\bar 1}
 \ket{S_{11'}} \ket{F}_{\bar 2} \,, \eqn\laim$$ 
and readily find that 
$$\bra{\omega_{12}} \Bigl(\,\int_D F^{[2]} \,\ket{F}\,\Bigr)_{1}\,\ket{\Psi}_2
  = (- 2\pi i) f(\T^2_1)  \,. \eqn\aim$$ 
On the other hand, the operator $\K$ can be thought as $\K = {1\over -2\pi i}
\int F^{[2]}$ acting from the right, and we would then write, using \newb\ 
$$\eqalign{
f(\K\B^1_1)  &= \bra{\omega_{12}}\Psi\rangle_1 \ket{F}_2\, {1\over (-2\pi i)}
\int F^{[2]}\,,\cr
& =  {1\over (-2\pi i)} \bra{\omega_{12}}\Bigl(\int F^{[2]}\ket{F}\Bigr)_1\, 
\ket{\Psi}_2  = f (\T^2_1)\,, \cr}\eqn\newbx$$
in agreement with our claim.  One can also verify directly 
that $f(\K\T^2_1)=0$ in agreement to what we would expect given that
$\K\T^2_1 = \K^2 \B^1_1 =0$. 

\section{String action and field equation for classical backgrounds}

Having justified the introduction of the $\B^1_1$, we now include this
space into the definition of the collection $\B^1$ by taking
$\B^1 = \B^1_1 + \B^2_2 + \cdots$. In this way 
the string action given in \sdes\ takes the simple form
$$S= Q + f(\B) \,, \eqn\fac$$
and the recursion relations \mnbv\ become 
$$\partial \B = \V'_3+ \K \B - \half  \{ \B , \B \} \,,\eqn\mnbvx$$ 
where we see that  the $(-\I\B)$ term in $\M\B$ appears now in  
the antibracket because of \decl, and the space $\T^2_1$ is now included
in $\K\B$ because of \nnont.  At this stage the recursion relations have 
become quite simple.\foot{It is tempting to simplify even more the 
recursion relations by
 absorbing the $\V'_3$ term in the right hand side. This
may be possible  by setting $\V'_3 = \K \V_2$, where $\V_2$ would be 
some formal space of two punctured spheres 
of dimension minus two. We will not explore this possibility here.
} The verification of consistency is now a trivial
computation. It is also manifest that the recursion relations
guarantee that the action satisfies the BV master equation. 
This implies that the action is gauge invariant.

Let us now examine the action in order to read the explicit
expressions for $\F$ and ${\cal Q}$.
The terms in the action  linear in the string field read
$$\eqalign{
S_{lin} &= f(\B^1_1) + \sum_{k=2}^\infty f(\B^k_1)  \,,\cr
&= \bra{\omega_{1\bar 1}} \Psi\rangle_1 \ket{F}_{\bar 1}
+ \sum_{k=2}^\infty 
{1\over k!} \int_{\B^k_1} \bra{\Omega^{[k]}_{1\bar 1 \cdots \bar k}}
\Psi\rangle_1 \ket{F}_{\bar 1} \cdots \ket{F}_{\bar k} \,. \cr } \eqn\lterms$$
By definition (Eqn.\act) 
$S_{lin} = \bra{\omega_{1\bar 1}} \Psi\rangle_1 \ket{\F}_{\bar 1}~$, 
and therefore we find
$$\ket{\F}_{1'} =  \ket{F}_{1'}
+ \sum_{k=2}^\infty 
{1\over k!} \int_{\B^k_1} \bra{\Omega^{[k]}_{1\bar 1 \cdots \bar k}}
 F\rangle_{\bar 1} \cdots \ket{F}_{\bar k} \ket{S_{11'}}\,. \eqn\nonl$$
The condition that $\ket{\F}$ vanishes is the equation that
selects a classical string background. This is a complicated nonlinear
equation for the state $\ket{F}$. While $\ket{F}=0$ is clearly
a solution, there may be solutions with $\ket{F} \not= 0$. The first
$\B$-space relevant to this equation is the space $\B^2_1$. This 
space satisfies $\partial \B^2_1 =\T^2_1 - \I\B^1_2$ and was constructed, 
except for the specification of the Weyl metric,
in Ref.[\zwiebachhms] sect.7.1.

The action also has terms quadratic in the string field, such terms
define the operator ${\cal Q}$ (see Eqn.\act). We can readily read
$${\cal Q} = Q + \sum_{k=1}^\infty 
{1\over k!} \int_{\B^k_2} \bra{\Omega^{[k]}_{12\hskip2pt\bar 1 \cdots \bar k}}
 F\rangle_{\bar 1} \cdots \ket{F}_{\bar k} \ket{S_{12'}}\,, \eqn\calq$$  
as an operator from the state space $2$ to the state space $2'$. 
We have therefore confirmed that the pair $({\cal Q} , \F )$ can
be constructed using the pair $(Q, F)$ and $\B$-spaces. By construction
we are guaranteed that the ``Bianchi identity" ${\cal Q}\ket{\F}=0$ holds.

\chapter{Comments and Open Questions}

In this paper we have sketched the construction of a string field
theory formulated around backgrounds that are not represented by
two-dimensional conformal field theories. The construction is
not complete in all its details. This should not be too surprising,
passing from  conformal
to non-conformal backgrounds is a major departure, and many issues arise.  

We have not been completely explicit about Weyl metrics on the
surfaces that make up $\B$ spaces.
While the minimal area problem [\zwiebachlong] determines 
consistent Weyl metrics on the surfaces defining the string 
vertices $\V$, we have not discussed a minimal area problem for surfaces
defining $\B$ spaces.
The required
 problem may be of the type proposed in [\zwiebachos] where one creates
interpolating moduli spaces by requiring 
that curves homotopic to the special punctures be longer than or equal to
a quantity $l$ that varies from zero to $2\pi$. 

Our analysis throughout this paper 
has been at the level of classical string theory.
We have  made no attempt to include the higher genus contributions to
the closed string action. No major difficulties are expected here, but
as usual, the vacuum contributions are probably quite subtle. 
The case of open strings has not been addressed.

Certainly the most important matter that has not been addressed 
is that of an explicit construction of the pair $(Q,F)$ satisfying
the postulated descent equations. Nor we have discussed the precise
construction of operator-valued forms on the moduli spaces of Riemann 
surfaces equipped with Weyl metrics. Moreover, since the $\B$ spaces
allow the antisymmetrized collision of special punctures, the 
generality of our construction hinges on the finiteness of the
antisymmetrized collision of $F$ states. We hope to address
these issues in Ref.[\zwiebachweyl].

We have obtained the field equation that characterizes a classical string
background. Such equation involved the spaces $\B^k_1$, spaces
with one ordinary puncture and any number of special punctures.
It is certainly of interest to understand the $\B^k_0$ spaces, that is,
the $\B$ spaces that have only special punctures. These spaces  contribute 
constants to the string action, constants that are essential for background
independence. Since the constants will be background dependent they
ought to give us insight into the function on theory space that represents
the string action. 

We have indicated in the table below the various string 
vertices with $k$ denoting the number of special punctures and $n$ the
number of ordinary punctures.  

$$\matrix{n: &0 & 1 & 2 & 3 & 4 &\cdots& k\cr
{}& {} & {} & {} & {} & \B^4_0 &\cdots& 4 \cr
{}& {} & {} & {} & \B^3_0 & \B^3_1&\cdots& 3 \cr
{}& {} & {} & \B^2_0 & \B^2_1 & \B^2_2 &\cdots& 2\cr
{}& {} & \B^1_0 & \B^1_1 & \B^1_2 & \B^1_3 & \cdots& 1\cr
{}& \V_0 & \V_1 & \V_2 & \V_3 & \V_4 &\cdots& 0\cr} $$
A few of the entries in this table are somewhat unclear. Conventionally
string vertices $\V_n$ exist only for $n\geq 3$. It may be useful to
define the space $\V_2$, as we mentioned in the last section. 
Probably $\V_0$ and $\V_1$ are really zero. Similarly while $\B^1_p$ spaces 
were formerly defined for $p\geq 2$ we have seen that $\B^1_1$ is
naturally defined and useful.  The space $\B^1_0$ may be zero.
It seems likely, however, that all $\B^k_0$ with $l\geq 2$ are non-vanishing.

Much was learned in the process of discovering how to write 
string field theory around conformal backgrounds. It is 
not unreasonable to believe that as much will be learned by
exploration of string field theory around arbitrary backgrounds.


\vfill
\break

\refout
\bye